%% file: str2.tex
\documentclass[prd,showpacs,preprintnumbers,floatfix,amsmath,amssymb,nofootinbib]{revtex4}
\usepackage{graphicx}

\newcommand{\ket}[1]{|#1 \rangle}           % ket-vector
\newcommand{\bra}[1]{\langle #1 |}          % bra-vector
\newcommand{\product}[2]{\langle #1 | #2 \rangle}   % scalar product
\newcommand{\mat}[4]{                   % 2x2 matrix
\left(
\begin{array}{rr}
#1 & #2 \\
#3 & #4
\end{array}
\right)
}
\newcommand{\inceps}[1]{\begin{array}{l} \includegraphics[height=18mm]{#1} \end{array}}
\newcommand{\smallpic}[1]{\begin{array}{l} \includegraphics[width=18mm]{#1} \end{array}}
\newcommand{\eq}[1]{(\ref{#1})}

\begin{document}

\title{Interactions of confining strings in SU(3) gluodynamics}

\author{V.\,G.~Bornyakov}
\affiliation{Institute for High Energy Physics, RU-142284 Protvino, Russia}
\author{P.\,Yu.~Boyko}
\author{M.\,N.~Chernodub}
\author{M.\,I.~Polikarpov}
\affiliation{Institute for Theoretical and  Experimental Physics,
B.Cheremushkinskaya 25, RU-117259, Moscow, Russia}

\begin{abstract}
We study the interaction energy of the
confining strings in the static rectangular tetraquark system in
SU(3) gluodynamics. Using a two-state approximation we calculate the energy
eigenvalues and corresponding eigenvectors of string states of different geometry.
The interactions are studied both for the co-aligned and for the counter-aligned parallel strings
as the functions of positions of the tetraquark constituents.
We formulate a simple soap-film model and show that with a good accuracy the
string interaction energy
corresponding to the ground state of the tetraquark system can be described by this soap-film model.
\end{abstract}


\preprint{ITEP-LAT/2005-11}

\maketitle
%===========================================================

\section{Introduction and Motivation}
\label{sect:intro}

The complicated phenomenology of the strong interactions at low energies
-- such as the spectrum of mesons and baryons as well as the properties of their resonances --
is determined by the forces between hadronic constituents, quarks and antiquarks.
These forces are dependent on spin orientations, relative
velocities and distances between the quarks and (anti-)quarks in the bound
states~\cite{ref:Reviews:QCDforces,ref:Reviews:correlators}.
The major role in the QCD spectrum is played by the quark confinement which implies
that the quarks are not observed as asymptotic states.

In SU(3) gluodynamics the confinement manifests itself as a strong
($\approx 1 \mbox{GeV}/\mbox{fm}$) attractive force between an external quark and an
anti-quark as they get separated by large enough distances.
It is believed that this force is a result of
a formation of a flux tube, or a QCD string, between the external color charges.
Since the QCD string has a constant energy per unit length the confining force is
asymptotically separation-independent.

In the real QCD the confining force is screened by pairs of quarks and
anti-quarks which emerge from the vacuum and break an exceedingly long QCD string.
Still the confining force caused by the QCD strings plays a crucial role in the
structure of hadronic bound states and their resonances
when these systems have sufficiently
large spatial size. It is very plausible that the hadron structure is influenced not
only by the interaction between the quarks themselves
({\it i.e.}, between the ends of the QCD
string) but also by the interaction between the segments of the QCD string(s) with
each other. The role of the QCD string interactions is the main subject of this
article.  Since the origin of the string and, consequently, the essence of
confinement, is encoded in the gluonic degrees of freedom, we restrict ourselves
to the quenched QCD case.

The three-quark (baryon) state represents a simplest non-trivial
case of the inter-string interactions. In the baryon there are two
possible geometries of the QCD string connecting three quarks: the
$\Delta$--like geometry which connects the quarks pairwise and the
$Y$--like geometry which connects quarks by three segments of the
string with a junction in a middle.
Lattice study of the baryon static potential revealed that the
potential agrees with the dominance of the two-body interaction
at small interquark separations~\cite{Takahashi:2002bw,Alexandrou:2002sn,Bornyakov:2004uv}.
This result is natural from the point of view point of the
asymptotic freedom which predicts that at short distances the
interquark interactions are reduced to the two-body interquark forces.

The $Y$-shape of the QCD string is difficult to observe at short distances due to the
dominance of the Coulomb potentials~\footnote{See, however, Ref.~\cite{ref:Ydominance}, where
an evidence of the $Y$-profile at short distances was found.}. However, as the interquark
separations get larger, the $Y$-ansatz behavior prevails
the two-body interactions~\cite{Takahashi:2002bw,Alexandrou:2002sn,Bornyakov:2004uv}.
The $Y$--type profile of the QCD string was also observed at large inter-quark separations in quenched and
unquenched lattice QCD in Refs.~\cite{ref:profile:DIK,Bornyakov:2004uv}
where Abelian projection was used as a useful tool.
In Ref.~\cite{Bissey:2005sk} the $Y$--type profile was observed in
unquenched lattice QCD with the help of gauge invariant operators.

The $Y$-geometry of the flux tube profile in baryon-like configuration
of quarks is consistent with the prediction~\cite{ref:Y-DGL} of
the dual superconductivity scenario~\cite{ref:DualSuperconductor},
which is assumed to work in the infrared region (for a review see
Refs.~\cite{ref:Reviews}). The field correlator method also predicts the
$Y$-geometry of the strings in the baryon~\cite{Kuzmenko:2003ck}.

In order to investigate the interaction between QCD strings it is convenient
to consider a geometry with two long disconnected segments of the string.
This corresponds to a generalization of the baryonic system to a system
consisting of four or more (anti)quarks. In this case, the interactions between
the QCD strings can be deduced from the interquark potentials. The multi-quark potentials
were intensively studied on the
lattice~\cite{ref:Ydominance,ref:Helsinki,ref:suganuma:4Q,ref:alexandrou:4Q,ref:suganuma:5Q,ref:alexandrou:5Q}.
In the simplest case of the SU(2) gauge theory, the chromoelectric string does not
have a directional dependence since quarks and anti-quarks in this theory are
indistinguishable from each other.
A numerical study of energies in the four-quark systems shown that the interaction,
or binding, energy is nonzero in the general case~\cite{ref:Helsinki}.
It was found  that if the four
quarks are partitioned into a colorless two-meson state with basic states
degenerate (or, almost degenerate) in energy,
then the interaction energy of the system is strongly enhanced. It was thus
shown that the energies of the four-quark states
cannot be simply represented as a sum in terms of two-quark potentials.
These features indicate that the mutual interaction of the chromoelectric strings
can not be ignored in the SU(2) four-quark systems.

In the SU(3) gluodynamics
investigation of the interaction energy of the tetraquark system with
two co-aligned mesons has shown~\cite{ref:suganuma:4Q,ref:alexandrou:4Q}
that the strings form a joint profile with two $Y$-type junctions
for small enough separations between the mesons.

The two-meson interactions were also studied analytically in the
dual superconductor model in the SU(2),
Ref.~\cite{ref:Suzuki:intermeson} and SU(3),
Ref.~\cite{ref:Y-DGL,ref:SU3:string:interaction} cases, as well as
in the gauge--invariant field--strength cumulant
approach~\cite{ref:Reviews:correlators,ref:Simonov:Shevchenko}.

Apart from purely theoretical interest in investigation of the inter-string
interactions (potentially applicable to physical systems) there is a direct experimental
motivation to study the 4Q systems in particular. Presently, candidates for a $4Q$ state include
the $X$(3872) resonance~\cite{ref:X} as discussed in
Ref.~\cite{ref:X:theory}, and $D_s$(2317) state~\cite{ref:Ds}. The knowledge of
the flux tube structure in the four quark systems seems to be important also
in hadron scattering reactions since these reactions may also involve a
rearrangement of the confining  strings.

In this paper we study the interactions of the QCD strings in the system consisting
of two quarks and two anti-quarks. The general considerations concerning the
interactions of the two (segments of)  confining
strings are discussed in Section~\ref{sect:general}. In order to calculate the
interaction energy we diagonalize the QCD Hamiltonian in the basis of the two
lowest energy states. As a guiding example, the calculation of matrix elements of
a Hamiltonian is done analytically in Section~\ref{sect:sc}
in a soap-film approximation which resembles the strong coupling expansion in the
lattice gluodynamics.  This calculation -- which is of a qualitative nature -- is
then compared with the results of the lattice simulations
of SU(3) gluodynamics discussed in Section~\ref{sect:data}. Our conclusions are
summarized in Section~\ref{sect:int}. The technical details of the numerical
procedures as well as principal output of simulations are presented in Appendices.

%-----------------------------------------------------------

\section{Interactions between QCD strings}
\label{sect:general}

The QCD strings have finite width but if they are long enough
they can be considered as line-like objects
connecting quarks or/and anti-quarks in three
spatial dimensions.
In a meson system the strings carry the chromoelectric flux
which originates at a quark and is absorbed at an anti-quark.
The interaction between the strings can be studied numerically with the help
of the correlator of two large Wilson loops separated by a large enough distance.
The correlator must be appropriately normalized by the expectation values of
the individual Wilson loops,
in order to subtract the energies of the infinitely separated quark--antiquark
pairs\footnote{Note that flipping the orientation of a pair, $\vec r\to - \vec r$
corresponds to the change $W \to W^\dagger$ of the corresponding Wilson loop.}.
The interaction energy $E_{int}(d,r)$ is then defined as
\begin{equation}
E_{\mathrm{int}}(d,r)  = - \lim_{t\to\infty} \frac{1}{t} \log
\frac{\langle W_{r\times t}(0) \, W_{r\times t}(d)\rangle}{
\langle W_{r\times t}(0)\rangle \, \langle W_{r\times t}(d)\rangle}\,,
\label{eq:E:int}
\end{equation}
where $W_{r\times t}$ is the rectangular Wilson loop of size $r \times t$,
$d$ is the distance between two parallel meson systems.
The Wilson loop $W_{r\times t}$ describes the evolution of the quark--antiquark
pair separated by the distance $r$ from time $t_0 = 0$ to time $t_1 = t$.

In Eq.~\eq{eq:E:int} the time extension $t$ of the Wilson loops should be taken
infinite. In order to study string interaction contribution into the full interaction
energy one should also take $r$ to be large in order to avoid effects of the short
range Coulomb interaction between static quarks. In the real lattice simulations
the above requirements are difficult to comply with.

\subsection{Flip-flop picture}

The important difference between the ``ideal'' strings discussed above and the
``real'' strings in lattice simulations is the fact that a lattice
string has a finite length $r$ and is evolved over a finite Euclidean time  $t$.
The constraints on $r$ and $t$ are due to the finiteness of the lattice volume.
As a result of this finiteness, there are some effects which make the
interpretation and realization of the Eq.~\eq{eq:E:int} difficult.

The finite $t$ implies the influence of higher excitations
(this effect is common to all lattice studies).
We will address the question of the $t$--dependence of our numerical results in
Appendix~\ref{app:time} where we show that the results are rather stable
against this kind of systematic errors.

If the distance $d$ is not large enough then the (anti)quark of one
quark-antiquark pair is interacting with the one of the other quark-antiquark
pair~\footnote{Note that the interactions of the quark constituents from the same
pair are already excluded by the proper normalization of
Eq.~\eq{eq:E:int}.} providing an unwanted contribution to the interaction energy
$E_{\mathrm{int}}$.
As a smearing technique we are using the hypercubic blocking (HYP) to
improve the signal to noise ratio for our observables, also
decreases effects of the short range inter quark interaction.
As it was shown in Ref.~\cite{hyp2}, the perturbative interaction -- which dominates
the interactions of quarks at small separations  -- is suppressed
by the HYP blocking. An interested reader can find the details of the HYP
blocking in Appendix~\ref{app:hyp}.

The effect of finite length $r$ is seen in a string rearrangement which is also known as
the "flip-flop" effect discussed in the context of SU(2),
Ref.~\cite{ref:Helsinki}, and SU(3), Ref.~\cite{ref:suganuma:4Q}, lattice gauge
theories. The flip-flop is a string rearrangement process
which happens as one changes the positions of the ends of the strings (quarks).
The rearrangement is caused by a requirement for the strings to
be in the ground state, i.e. in a state with lowest possible (for given boundary
conditions) value of energy. As we will see later this is basically the same as
the requirement for the string configuration to have a shortest possible length.

\begin{figure}[!htb]
\includegraphics[width=60mm]{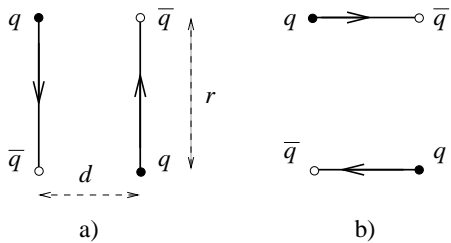}
\caption{A schematic illustration of the flip-flop picture: (a) the longitudinal and (b) the transverse string configurations.}
\label{fig:1}
\end{figure}
\begin{figure}[t]
\includegraphics[width=60mm]{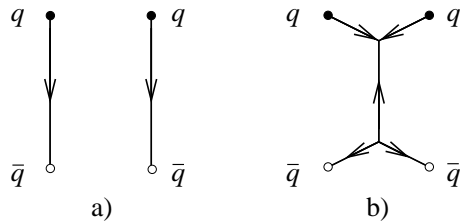}
\caption{The string configurations for the quark pairs with the same orientation: (a) the longitudinal configuration;
(b) The transverse configuration with two string junctions. The minimal length corresponding to the case (b) is $L_b=r+\sqrt{3}d$.}
\label{fig:2}
\end{figure}

The rearrangement is illustrated in Fig.~\ref{fig:1} for the
simplest case of two parallel strings with the same length and opposite
orientations. This string configuration is called below the ``longitudinal''
configuration.
The two quarks and two antiquarks shown in Fig.~\ref{fig:1}(a) can also be
connected by strings in another way shown schematically in Fig.~\ref{fig:1}(b).
Below we call the latter ``transverse'' configuration. Let us keep the size $r$
of the pair fixed and decrease the separation $d$. When $d$ becomes smaller
than some critical value  $d_c$ the energy of the transverse configuration becomes smaller than the one for
the longitudinal configuration. It is energetically favorable for the strings to realign at this point.
Having neglected an interaction between the segments of the string, we can
estimate an energy of the string configuration as $E \approx \sigma
L$, where  $L$ is a total length of the string configuration, and $\sigma$ is the
universal string tension\footnote{The universality of the string
tension means that the string tensions in the heavy mesons and baryons are the
same. This universality is supported by the numerical evidence obtained
in lattice studies of the static  baryon~\cite{Takahashi:2002bw}.}. Obviously, due
to a simple geometry of the string configuration shown in Fig.~\ref{fig:1},
the distances $d_c$ and $r$ are approximately equal in this particular example. However,
in other systems the competitive configurations can be more complicated, like the one shown in Fig.~\ref{fig:2}
for the case of equally oriented strings. The critical separation in this case
is estimated to be $d_c \simeq r/\sqrt{3}$.

If one treats the string configuration classically, the string
re-alignment indicates existence at the flip-flop point $d = d_c$
of (at least) two degenerate energy minima in the space of the
string configurations. On the quantum level, the degeneracy leads
to the level splitting effect, as it is well known from the quantum
mechanics. The level splitting is caused by the tunneling
transitions between the wells of potential energy. Therefore one
expects that the ``transverse'' and the ``longitudinal'' string
states are mixed with each other due to the tunneling transitions
and, consequently, the degeneracy is taken away. Technically, the
tunneling transitions lead to an appearance of the non-diagonal
matrix elements of the Hamiltonian in the basis of the original
states ({i.e.}, the states corresponding to the far-separated
"unperturbed" individual wells). To get the correct energy states
one needs to diagonalize the Hamiltonian taking into account the
tunneling transitions. The true ground state of the string in a tetraquark
is a weighted mix of all "unperturbed" states of this system. In the two-- and three-quark
systems a similar investigation was performed, respectively, in Refs.~\cite{ref:Morningstar} and
\cite{ref:gluonic:excitation} in order to investigate gluonic excitations.

In order to treat the degeneracy of the ground state exactly one needs to consider
not only the ground state itself but also one should know all the
states of the system. However, the higher is the state the lower is its mixing
with the ground state. Therefore, one usually considers the two-state approximation,
which takes into account only the ground state and the next energy state of the
string system. Technically, this approach corresponds to the
diagonalization of the Hamiltonian in the two dimensional subspace of the states.

%-------------------------------------------------
\subsection{Two-level Approximation}

We can construct ("prepare") an operator to study the  particular
configuration of the quarks and anti-quarks by connecting the corresponding sets
of points in a three-dimensional time-slice by gauge transporters. These transporters,
called sometimes as "Schwinger lines", are the open Wilson lines
in the fundamental representation of the gauge group. Such a set of lines
transforms under gauge transformations as a set of quarks and anti-quarks.
The better is the overlap of this operator with the real multiquark ground state
(determined by the QCD dynamics) the smaller are effects of higher excitations.
In order to improve this overlap we use the APE smeared gauge transporters, see
Appendix~\ref{app:ape} for details.

We consider the configurations of the Wilson lines shown in Figures~\ref{fig:1} and \ref{fig:2}.
The corresponding states define a two dimensional subspace of the state space discussed in the previous
Subsection. The explicit definition of the basis of the states depends on
positions of quarks and anti-quarks. We consider two simple geometries: co-aligned (Fig.\ref{fig:2}) and oppositely aligned (Fig.\ref{fig:1})
parallel strings of the same length, and we present the exact expressions concurrently where the relative orientation makes a difference.

The initial states of our basis -- corresponding to the classical string configurations -- are shown in Table~\ref{tab:bdef}. The oriented
lines mean the gauge transporters, the closed contour is the Wilson loop, and $N_c=3$ is the number of colors used to normalize the initial
states\footnote{Note that the number of the string segments coming to the string junction is equal to the number of colors.}.
We refer to these basis states as $\ket{A}$ and $\ket{B}$ regardless of the
orientation.  Although the states are normalized, $\product{A}{A} =
\product{B}{B} = 1$, they are not, in general, orthogonal to each other,
\begin{equation}
\label{eq:gdef}
g \equiv \product{A}{B} \neq 0\,.
\end{equation}
The basis states overlap $g$ is proportional to the spatial Wilson loop and also shown in Table~\ref{tab:bdef}.

\begin{table}[!ht]
\begin{tabular}{l|r|r}
\hline \hline
& Opposite orientation & Same orientation \\
\hline
    $\ket{A}$ &
    $N^{-1}_c \smallpic{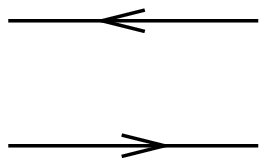} \ket{0}$ &
    $N^{-1}_c \smallpic{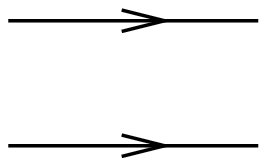} \ket{0}$
\\
    $\ket{B}$ &
    $N^{-1}_c \smallpic{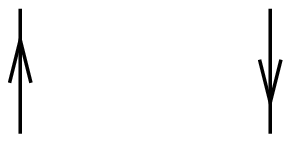} \ket{0}$ &
    $N^{-1/2}_c \smallpic{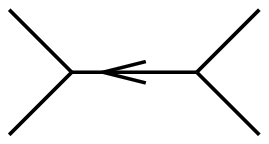} \ket{0} $
\\
    $\product{B}{A}$ &
    $N^{-1}_c \bra{0} \smallpic{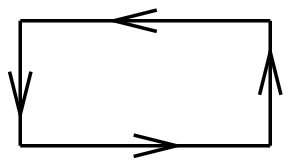} \ket{0}$ &
    $N^{-1/2}_c \bra{0} \smallpic{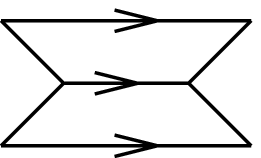} \ket{0} $
\\
\hline \hline
\end{tabular}
\caption{The graphical definition of the states from the initial basis and of their overlap.}
\label{tab:bdef}
\end{table}

Let us consider the diagonalization of the Hamiltonian $\hat{H}$ in the space $\ket{A} \oplus \ket{B}$. To this end
one can first diagonalize the Euclidean time evolution operator
\begin{equation}
\label{eq:sh}
\hat{S}(t) = \exp(-\hat{H} t)\,,
\end{equation}
the matrix elements of which can be calculated in lattice simulations.
Despite the operator identity (\ref{eq:sh}) is not correct in the finite basis, $S_{ab} \neq \exp(-H_{ab}t)$,
one expects that only the lowest energy states contribute to the evolution operator matrix elements for
large enough evolution time $t$. We will see that the $t$-dependence of the energy levels,
defined via evolution operator eigenstates, has a form of $1/t$ corrections and, consequently, can be neglected for large enough $t$.

The main observables calculated in our numerical simulations are the $S$ matrix elements:
\begin{equation}
\label{eq:sdef}
{\mathbf S}(t) =\mat{\bra{A} e^{-Ht} \ket{A}}   {\bra{B} e^{-H t} \ket{A}}
        {\bra{A} e^{-Ht} \ket{B}}   {\bra{B} e^{-Ht} \ket{B}}\,.
\end{equation}
They are expressed as the Wilson loops correlators which for the case of the oppositely aligned strings can graphically be represented as follows:
\begin{eqnarray}
\label{eq:smat1}
&{\mathbf S}(t) =
&\bra{0} \mat
{\inceps{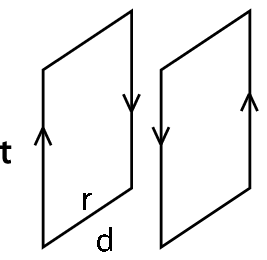}}
{N^{-1}_c \inceps{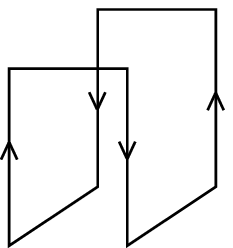}}
{N^{-1}_c \inceps{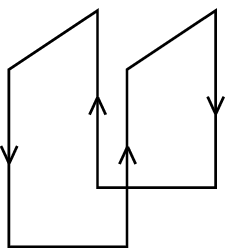}}
{\inceps{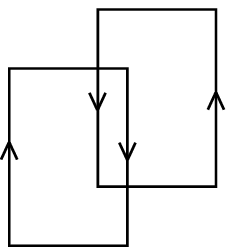}}
\ket{0}\,.
\end{eqnarray}
For co-aligned strings the graphical representation is a bit more complicated:
\begin{eqnarray}
\label{eq:smat2}
&{\mathbf S}(t) =
&\bra{0}
\mat{\inceps{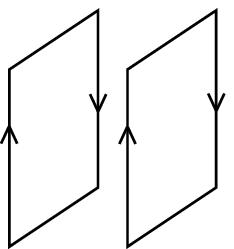}}  {N^{-1/2}_c \inceps{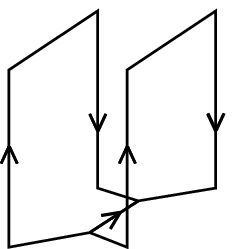}}  {N^{-1/2}_c \inceps{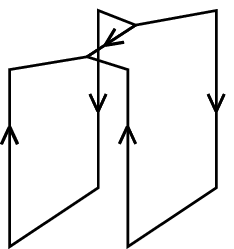}}
{\inceps{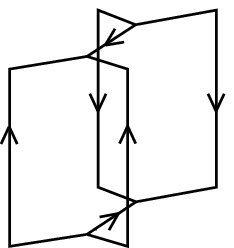}}
\ket{0}\,.
\end{eqnarray}

The problem to find eigenfunctions and eigenvalues of the matrix $S$
\begin{equation}
\hat{S}(t)\ket{E} = \exp(-Et) \ket{E}\,,
\end{equation}
can be reformulated as a generalized eigenproblem in terms of the matrix elements,
\begin{equation}
\label{eq:eigen}
S_{ab}(t) h_b = \lambda(t) T_{ab} h_b\,,
\end{equation}
where $a,b \in \{A,B\}$ and $\ket{E} = h_A \ket{A} + h_B \ket{B}$. The matrix
${\mathbf T}$ in the right-hand side of Eq.~(\ref{eq:eigen}) is nothing but
the overlap of the basis states, discussed above,
\begin{equation}
\label{eq:s0def}
{\mathbf T} = \mat{1}{g}{g}{1}\,.
\end{equation}

Equation (\ref{eq:eigen}) together with the normalization condition
\begin{equation}
\product{E}{E} = h_A^2 + h_B^2 + 2 g h_A h_B = 1
\end{equation}
defines two eigenstates $\ket{E_{0,1}}$ of the
evolution operator (and, with all precautions, of the Hamiltonian).
These eigenstates correspond to the energies
\begin{equation}
\label{eq:edef}
E_{0,1} = -\frac{\ln(\lambda_{0,1})}{t}\,.
\end{equation}
In order to solve the eigenproblem one should know all the matrix elements $S_{ab}(t)$. To understand the qualitative properties of the
eigenvalues we perform an analytical calculation in the soap-film approximation in the next Section.

Note that we call the energy $E_0$ as the ground state energy while the energy $E_1$ is refereed to as the excited energy. This standard terminology
-- the ground state corresponds to the lowest energy while the excited state to the higher one -- should not be taken literally. Indeed, the states
correspond to different string configurations in the sense of geometry
(like in Figures~\ref{fig:1} and \ref{fig:2}), and they most
probably should not be interpreted as the excited ({\it i.e.}, vibrational) modes of one of (the segments of) the strings.
Such excited modes were studied in the case of a single QCD string in a static meson~\cite{ref:Morningstar}, and in a case
of the simply-connected $Y$-shaped string segments in the case of a static baryon~\cite{ref:gluonic:excitation}. In the sense
of such excited mode both energies $E_0$ and $E_1$ correspond to the ground states of the particular string configurations
visualized schematically in Figures~\ref{fig:1} and \ref{fig:2}.

%-----------------------------------------------------------
\section{Soap-film approximation}
\label{sect:sc}
\begin{table*}[!t]
\begin{tabular}{l|r|r}
\hline \hline
 & Opposite orientation & Same orientation \\
 \hline
$S_{AA}$ &
    $~~~\inceps{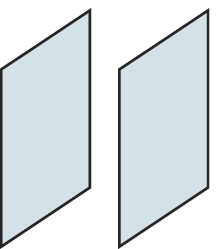}~+~\displaystyle{\frac{1}{N^2_c}} \inceps{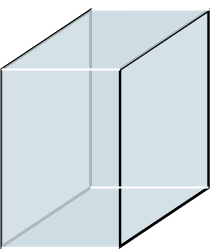}$  &
    $~~~\inceps{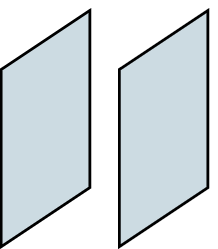}~+~\displaystyle{\frac{1}{N_c}} \inceps{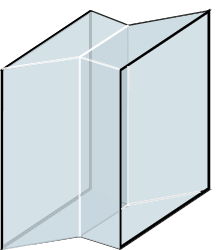}$  \\
$S_{BB}$ &
    $~~~\inceps{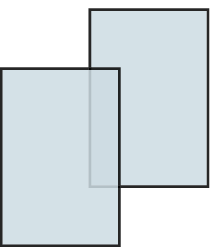}~+~\displaystyle{\frac{1}{N^2_c}} \inceps{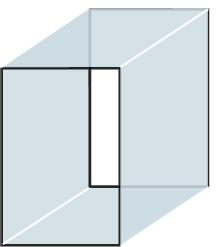} $ &
    $~~~\inceps{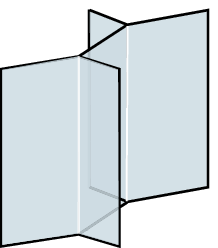}~+~\displaystyle{\frac{1}{N_c}} \inceps{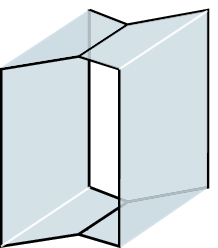}$ \\
$S_{AB}$ &
    $~~~\displaystyle{\frac{1}{N_c}}
        \left( \inceps{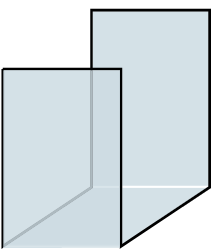}~+~\inceps{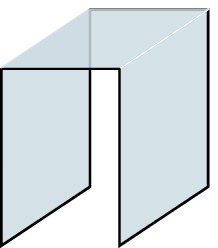} \right)$  &
    $~~~\displaystyle{\frac{1}{\sqrt{N_c}}}
        \left( \inceps{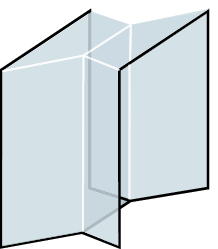}~+~\inceps{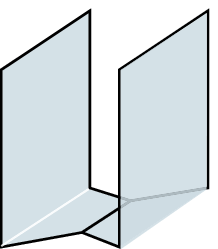} \right)$ \\
\hline \hline
\end{tabular}
\caption{\label{tab:soap}
The soap-film expansion of the evolution matrix elements ($S_{BA} = S_{AB}$).
}
\end{table*}

In the soap-film approximation we evaluate the average of the Wilson loops
taking into account only the area contributions
\begin{equation}
\label{eq:scdef}
\bra{0} W_1 \dots W_n \ket{0} \Longrightarrow \sum_i k_i \exp(-\sigma S_i),
\end{equation}
where the sum is taken over all topologically distinct and locally minimal surfaces with the area
$S_i$ which are spanned on a given set of oriented contours.
The integer pre-factors $k_i$ count the number of string configurations which can be spanned on the given set of loops.
Finally, $\sigma$ is the universal string tension.

In general, Eq.~\eq{eq:scdef} can be justified either in the
strong coupling regime (discussed in Ref.~\cite{ref:alexandrou:4Q}
for the multi-quark interaction) or in the infrared region, where
the perimeter-like contributions to the Wilson loops can be
neglected. Moreover, the soap-film expansion -- applied in the
infrared region -- coincides with the field-strength correlator
expansion performed in Ref.~\cite{ref:Simonov:Shevchenko} to study
interactions of the Wilson loops in the Gaussian approximation.

The essential point is that in Eq.~\eq{eq:scdef} we neglect the string fluctuations, the contribution of which is small compared with the leading
area term. Note that in the discussed approximation~\eq{eq:scdef} any
{\it local} string interactions ({\it e.g.} via exchange of a particle) are neglected.
On the other hand, the effect of re-alignment is still encoded in Eq.~\eq{eq:scdef}, therefore this representation ideally suits our needs as
a simple theoretical model.

The expansion of the matrix elements (\ref{eq:smat1}),(\ref{eq:smat2}) is shown in Table~\ref{tab:soap}. Overall normalization of the matrix elements
can easily be obtained from the normalization of the basis states. The color factors $1/N_c$ and $1/N^2_c$ in the non-planar contributions to
the diagonal matrix elements appears in the strong coupling expansion and in the field-strength approach~\cite{ref:Simonov:Shevchenko}.

Decomposing the surfaces into the spatial and the time-like faces,
\begin{equation}
\label{eq:faces}
\exp(-\sigma S) = \prod_{faces} \exp(-\sigma S_i)\,,
\end{equation}
and identifying the spatial contributions with the basis states overlap $g$, one derives
\begin{equation}
{\mathbf S}(t) =
f_r \mat{1}{g}{g}{g^2} + f_d \mat{g^2}{g}{g}{1}\,,
\end{equation}
where
\begin{equation}
\label{eq:fr}
f_r = \exp(-2\sigma r t)\,,
\end{equation}
and
\begin{equation}
\label{eq:fd}
f_d = \left\{
\begin{array}{lr}
\exp(-2\sigma d t) & \textrm{opposite orientation} \\
\exp(-\sigma(\sqrt{3}d + r) t) & \textrm{same orientation}
\end{array} \right.\,.
\end{equation}

The characteristic equation corresponding to Eq.~(\ref{eq:eigen}),
\begin{equation}
\det({\mathbf S}(t) - \lambda {\mathbf T}) = 0\,,
\end{equation}
where ${\mathbf{T}}$ is defined in (\ref{eq:s0def}), has the following roots
\begin{equation}
\label{eq:lroots}
\lambda_{0,1}~=~\frac{1}{2}(f_d + f_r \pm \sqrt{(f_d - f_r)^2 + 4g^2 f_d f_r})\,.
\end{equation}
Finally, the energy levels are obtained with the help of Eq.~(\ref{eq:edef}).

We are interested in the two regions of the parameter space: (i) the region near the flip-flop point $r = d$ and (ii) the ``interaction'' region
corresponding to the two relatively long strings placed near each other, $d \ll r$.

\begin{itemize}
\item[$d = d_c$] At the flip-flop point $f_r = f_d = f$. Then Eq.(\ref{eq:lroots}) becomes
\begin{equation}
\lambda_{0,1}|_{d=d_c} = f(1 \pm g)\,,
\end{equation}
giving the levels splitting (valid for both string orientations)
\begin{equation}
(E_1-E_0)|_{d=d_c} = \frac{1}{t} \ln \frac{1+g}{1-g}
= \frac{2g + o(g)}{t}\,.
\label{split2}
\end{equation}
Thus the level splitting is just a finite $t$ effect in the case when the basis of the Hamiltonian eigenstates is truncated
(the true Hamiltonian eigenvalues do not depend on $t$ in the full basis).

\item[$d \ll r$] At the interaction region the energy splitting can be obtained by an expansion of the square root in Eq.(\ref{eq:lroots})
in powers of $f_r/f_d$,
\begin{subequations}
\begin{eqnarray}
\lambda_0 &=& f_d\{1 + f_r/f_d \cdot g^2 + o(f_r/f_d)\} \\
\lambda_1 &=& f_r\{1-g^2- f_r/f_d \cdot g^2(1-g^2) + o((f_r/f_d)^2)\}\,.
\end{eqnarray}
\end{subequations}
Then the leading contribution to the energy levels takes the following form (the orientation is explicitly indicated),
\begin{subequations}
\begin{eqnarray}
E_0 |_{d \to 0} &=& \left\{
\begin{array}{ll}
2\sigma d &  \textrm{opposite} \\
\sigma(\sqrt{3}d + r) & \textrm{same}
\end{array} \right. \\
E_1 |_{d \to 0} &=& 2\sigma r - \frac{\ln (1-g^2)}{t} = 2\sigma r + \frac{g^2 + o(g^2)}{t}\,,
\end{eqnarray}
\end{subequations}
leading just to the ``classical'' string configurations up to the
$1/t$ corrections. In this case the interaction is absent as it was expected.
\end{itemize}

An exact (unexpanded) expressions for the eigenvalues as functions of the
parameters $d$ and $r$ are too bulky we skip their functional forms
here. However, we present the soap-film
results in the graphical form
in the next Section in order to compare them with the lattice data.

For the comparison of the lattice data and the soap-film predictions, the latter
should be
improved to cover the small $d$ and $r$ regions in a parameter space, for
which the Wilson loops are not saturated by area law. A possible way to do this systematically
is to make in Eq.~(\ref{eq:faces}) a substitution of single Wilson loop averages for time-like faces
(which is supposed to be an area-law behavior in the soap-film formalism) by the expression
\begin{equation}
\exp(-\sigma S_i) \Rightarrow \langle W_i \rangle = \exp(-V(r_i) t)
\end{equation}
where $i$ runs over all plain time-like faces of size $r_i \times t$, bounded by rectangular
Wilson loop and $\langle W \rangle$ is related to the static quark-antiquark potential $V(r)$.
In fact we already performed this trick when we used for the spatial faces
the overlap $g(r,d)$ instead of the area-law expression which would originate from the soap-film formalism.

The equations (\ref{eq:fr}),(\ref{eq:fd}) now look as:
\begin{equation}
\label{eq:fr:new}
f_r = \exp(-2V(r)t),
\end{equation}
and
\begin{equation}
\label{eq:fd:new}
f_d = \left\{
\begin{array}{ll}
\exp(-2V(d) t) & \textrm{opposite orientation} \\
\exp(-(4V(d/\sqrt{3}) + V(r - d/\sqrt{3})) t) & \textrm{same orientation}
\end{array} \right.\,.
\end{equation}
All other expressions remain unchanged.
In order to perform the improvement we evaluate numerically the static potential $V(r)$ and the overlap function $g(r,d)$.
Below we use the improved soap-film formulation instead of the naive area-law expansion.

%-----------------------------------------------------------

\section{Lattice Data}
\label{sect:data}
%===========================================================
\begin{figure*}
\input{eps/energy.opp.tex} \hfill \input{eps/energy.same.tex}
\caption{\label{fig:energy} The energy eigenvalues $E_{0,1}(r,d)$ as the functions
of the string separation $d$ at the fixed string length $r=0.84$\,fm. The left (right) plot corresponds to the oppositely aligned
(co-aligned) strings. The points corresponds to the numerical data obtained in simulations while the lines are
the soap-film predictions. The statistical errors of both data points are of the order of the symbol size.
The error in estimation of the soap film prediction is of the same magnitude. Note that in all
our numerical results the perturbative contributions are suppressed by the HYP procedure}
\end{figure*}
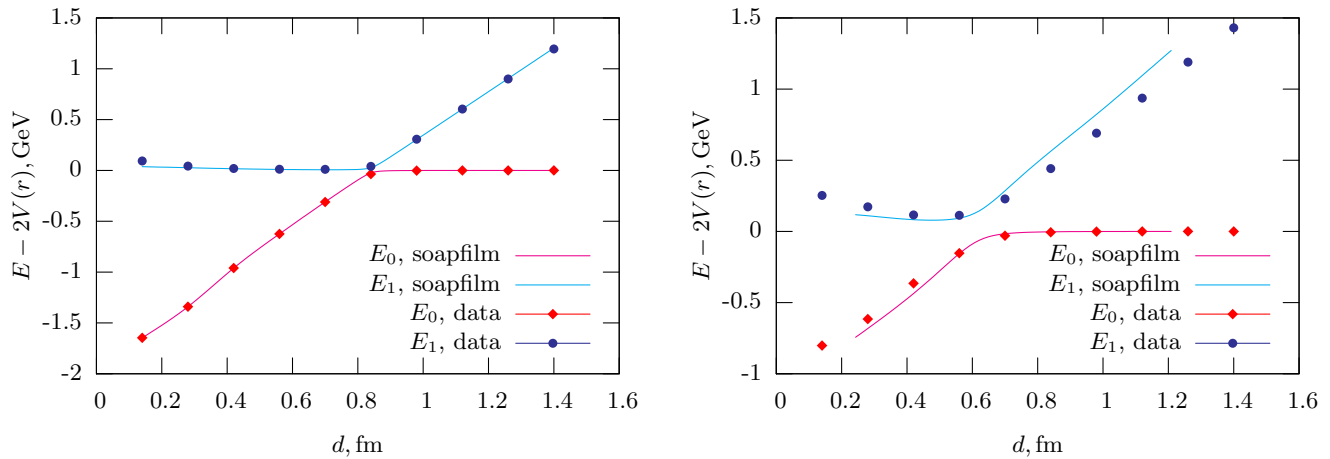
%-----------------------------------------------------------
\begin{figure*}
\input{eps/ed.0.tex} \hfill \input{eps/ed.1.tex} \\
\centerline{Ground state \hspace{7cm} Excited state}
\caption{\label{fig:ediff} The difference between $\delta E = E-E_{\mathrm{soap}}$ the measured energy levels and the prediction of the soap-film
approximation for the  oppositely aligned strings of the length $r = 0.84$\,fm. The points represent $\delta E$ while the band represent the
statistical errors of soap-film prediction. Note the MeV scale of vertical axis
compared to the GeV scale of interaction energy in Fig.~\ref{fig:energy}.}
%-----------------------------------------------------------
\input{eps/ed.3.tex} \hfill \input{eps/ed.4.tex} \\
\centerline{Ground state \hspace{7cm} Excited state}
\caption{\label{fig:ediff:same} The same as in Fig.~\ref{fig:ediff} but for the likewise oriented strings.}
\end{figure*}
%===========================================================

In order to obtain energy eigenvalues and eigenvectors one needs to know the evolution matrix elements $S_{ab}$ given in
Eq.(\ref{eq:smat1}),(\ref{eq:smat2}), the overlap of the basis states $g(r,d)$ and the static potential $V(r)$. Since all mentioned observables are
expressed in terms of the Wilson loops correlators, we perform a lattice computation which is straightforward. The details of numerical setup and
computation details are given in Appendix~\ref{app:num}.

The analysis is performed similarly to the analytical case of the
soap-film expansion presented in the previous Section. The energy
levels of oppositely aligned and co-aligned strings are provided
as functions of parameters $r,d$ in, respectively, Tables
\ref{tab:e.opp} and \ref{tab:e.same} of Appendix~\ref{app:numerical}.

In Fig.~\ref{fig:energy} we show the energy eigenvalues
$E_{0,1}(r,d)$ as the functions of the string separation $d$ for relatively large
length of the strings, $r = 6a = 0.84$\,fm. In these figures the energy of
the two strings $2V(r)$ is subtracted to focus on the interaction between the strings (this subtraction corresponds to
the normalization factor in the denominator of Eq.~\eq{eq:E:int}). Both figures clearly show the string re-arrangement
at some point. We found that the gap between the states in the flip-flop points exists in all the
studied cases and it is of the same order for the co-aligned and oppositely aligned
strings.  The soap film prediction is shown in Fig.~\ref{fig:energy} by solid
lines\footnote{Note that the soap-film inputs (the overlap $g$ and the static
potential $V$) were calculated for the on-axis
values of parameters, which correspond to integer-valued distances in units of
the lattice spacing $a$.  In order to achieve a more frequent sampling of the
soap-film predictions, the values and the statistical errors of $g(r,d)$ and
$V(r)$ were interpolated to the non-integer $r/a$ and $d/a$ with the step $a/10$.
The ambiguity of such interpolation does not influence the comparison of the
soap-film with the data, since the data itself was calculated only for on-axis
parameters.}.

The interaction energies -- shown in Fig.~\ref{fig:energy} -- are in a very good agreement with the soap-film
prediction discussed in previous Section. To compare the soap-film model with the
numerical data we plot the difference between the predicted and measured energies for the
oppositely (likewise) aligned strings of the length $r = 0.84$\,fm in Fig.~\ref{fig:ediff} (Fig.~\ref{fig:ediff:same}).
One can clearly see that in the case of the opposite orientations (Fig. \ref{fig:ediff}) the
deviation is very small (of order of $10$\,MeV) compared to the strings energies
(approximately
$1$\,GeV) and is enhanced in the flip-flop point. The deviation is -10 MeV at the flip-flop point
both for the ground state and for most length of the string in the exited state.
At the lowest distances the deviation is of the order of 50 MeV for the exited
state. This might be due to overlap of the cores of the strings at these distances.

Besides the gap in the re-arrangement point, the energy of the ground state of the strings configuration with the
same orientation deviates from the soap-film prediction at the string separations smaller than
$d_c \simeq 0.4$\,fm, Fig.~\ref{fig:ediff:same}. The exited state has a bit different slope
than the soap-film prediction. The deviations are of order of 100 MeV which is still at 10\%
level compared to the energy of a string.
The 10\% deviations observed in Fig.~\ref{fig:ediff:same} can be explained by
imperfection of the initial and final states of the string. Indeed, the initial string is made of segments
of the Schwinger lines, the length of which is of order of the physical string width.

\begin{figure*}[!hbtp]
\input{eps/split.t4.tex} \hfill \input{eps/split.t6.tex}
\caption{\label{fig:split}
Ground and excited energies for oppositely aligned strings at the flip-flop point
vs. the distance $r$ between the strings for $t=4a$ (left) and $t=6a$ (right).
The curves show the soap-film model result.}
\end{figure*}
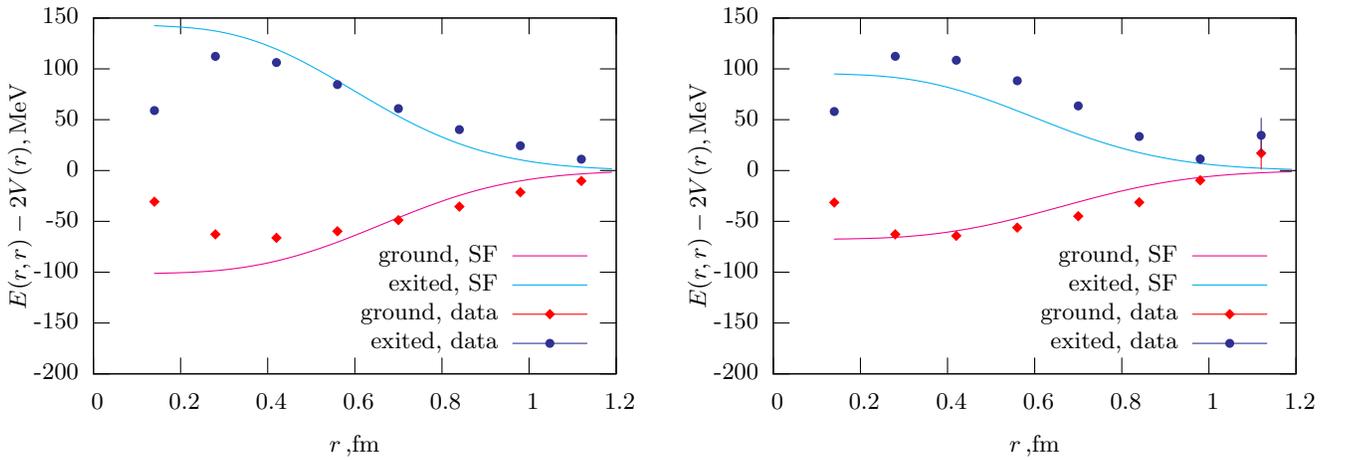

In Fig.~\ref{fig:split} we show the energies $E_{0,1}(r,d)$ as
functions of $r$ for $d=d_c$. One can see that at the flip-flop
point the SF model works very nice for all distances $r$ apart
from small ones. The discrepancy at small $r$ might be due to
effects of HYP procedure. Comparison of data obtained with $t=4$
and $t=6$ Wilson loops show that for large enough $r$ the level
splitting indeed decreases with increasing $t$, as predicted by
Eq.~\ref{split2}. From our data we can not conclude whether the
splitting goes to zero in the limit $t \to \infty$.

Let us stress that the accuracy of the soap-film expansion is amusingly high taking into account the
scales of the axis in Figures~\ref{fig:ediff} and \ref{fig:ediff:same} and the fact
that {\it no fits} are made to achieve this nice agreement between the soap-film
predictions and the numerical results.

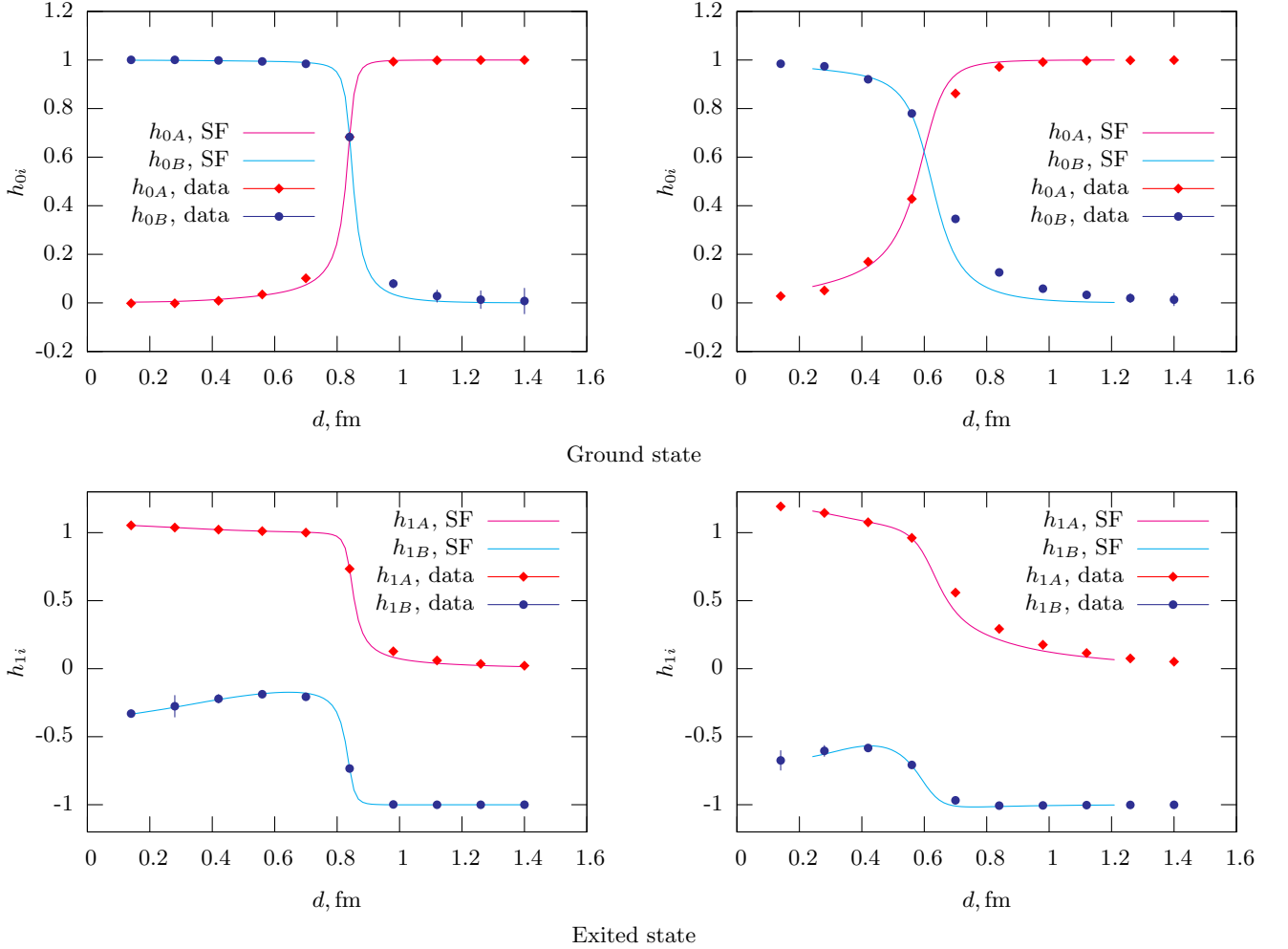
\begin{figure*}
\input{eps/h0.opp.tex} \hfill \input{eps/h0.same.tex} \\
Ground state \\
\input{eps/h1.opp.tex} \hfill \input{eps/h1.same.tex} \\
Exited state
\caption{\label{fig:eigenstates} The same as in Fig.~\ref{fig:energy} but for the
expansion of the energy eigenstates in the basis of the initial states $|A\rangle$
and $|B\rangle$. The left (right) plot corresponds to the oppositely
(co-aligned) aligned strings while top (bottom) row corresponds to ground (excited) states.}
\end{figure*}

In addition to the energy we study the eigenstates of the Hamiltonian. The energy
eigenstates $\ket{h_{0,1}}$ are shown as the functions of $d$ for
fixed $r=0.84$\,fm in Fig.~\ref{fig:eigenstates}. One can observe the flip-flop
behavior in the coefficients $h_{iA}$ and $h_{iB}$,
\begin{equation}
|h_i\rangle = h_{iA}(r,d) |A\rangle +h_{iB}(r,d)|B \rangle\,,
\label{eq:h}
\end{equation}
which describe the expansion of the ground and excited states in terms of the original states $|A\rangle$ and $|B\rangle$. Note that the
re-arrangement of the strings is visible simultaneously both for the ground and for the excited states. The absolute value of the coefficients may be
larger than unity since the states  $|A\rangle$ and $|B\rangle$ are not orthogonal to each other.

%====================================================
%====================================================

The soap-film prediction is also shown in Fig.~\ref{fig:eigenstates}. As in the case of the energy levels,
it is almost perfectly consistent with the data.

\section{Interpretation and Conclusions}
\label{sect:int}

We studied the interaction energy of the tetraquark system focusing on the interaction between the string states
which appear in this quark system. Using a two-state approximation we
calculate the eigenfunctions and eigenvalues
corresponding to the string states of different geometries as the
function of the positions of the tetraquark constituents. The data is in a good agreement with the prediction of the simple soap-film model which
can be considered as an analog of the strong coupling expansion or as a zero order (Gaussian) approximation in the field strength correlation
method~\cite{ref:Reviews:correlators,ref:Simonov:Shevchenko}. We observe a small deviation from the soap-film prediction at the flip-flop positions
of the tetraquark constituents where the classical energies of the different string configurations (allowed by the (anti-)quark arrangement) coincide.
This deviation is very small (of the order of 10~MeV for the ground state) compared to the string
energies (which are typically of the order of 1~GeV). This means that
in the realignment point a possible ``additional'' ({\it i.e.}, is not caused by the flip-flop) interaction between
the string segments is very small. Our conclusion is in agreement with similar observation made in the SU(2) gauge theory~\cite{ref:Helsinki}.

It is interesting to discuss the relation of our results to the
dual superconductor model in QCD which also predicts a certain
interaction between the QCD strings. The dual superconductor model
was invented to describe the confining properties of the
Yang--Mills vacuum. The main role in the dual superconductor is
played by special configurations of the gluonic field which are
called Abelian monopoles. Such a configuration has a Dirac-like
monopole singularity in its center after a suitable Abelian gauge
is fixed. The importance of the monopoles is stressed by the
observation of the Abelian dominance~\cite{ref:AbelianDominance}
implying that the tension of the chromoelectric string is dominated by the
Abelian monopole contributions. The model predicts that the QCD
strings are interacting via exchange of the dual gauge boson and
the dual Higgs fields.

Various lattice simulations indicate that the non-Abelian gauge
theories (both SU(2) and SU(3)) correspond to the border between
type-1 and type-2 dual superconductors. At this border the masses
of the both dual fields coincide and the parallel strings with the
same orientation are not interacting at all while the string with
opposite interactions are always attracting. This interaction goes
beyond the soap-film prediction which works very well as we have
just noted. Moreover, we see that even at smallest distances the
deviation of the ground state energy from the soap-film picture is
very small, and therefore the traces of the additional
inter-string interaction are not found.

On the other hand it is expected that the masses of both fields
are of the order 1~GeV which, in turn, is the scale of the lattice
spacing used in our model. Since the inverse mass of the dual
fields is of the same order as the lattice spacing, our data can
not either confirm nor reject the validity of the applicability of
the dual superconductor approach as a possible model for the
string-string interactions in QCD.

%-----------------------------------------------------------
\begin{acknowledgments}
The authors are grateful to Fedor Gubarev and Vladimir Shevchenko for interesting discussions,
and to Hideo Suganuma for valuable comments on the content of the paper.
This work is partially supported by grants
RFBR 05-02-17642,
RFBR 05-02-16306,
RFBR 04-02-16079,
RFBR-DFG 03-02-04016,
DFG-RFBR 436 RUS 113/739/0,
and by the EU Integrated Infrastructure Initiative Hadron Physics (I3HP)
under contract RII3-CT-2004-506078. P.B. is supported by the Dynasty Foundation Fellowship.
\end{acknowledgments}

%-----------------------------------------------------------
\appendix

\section{Numerical Setup}
\label{app:num}
We use a set of quenched $SU(3)$ gauge field configurations generated by MILC~\cite{MILC} from the Gauge
Connection~\cite{GC}. The parameters of these configurations are shown in Table~\ref{tab:param}.
To enhance statistics the combination of spatial APE smearing and time-like HYP smearing was used.
The corresponding definitions and parameters are given in Appendix~\ref{app:ape} for the APE smearing and
in Appendix~\ref{app:hyp} for the HYP smearing.
\begin{table}[!ht]
\begin{center}
\begin{tabular}{ll}
\hline \hline
Action:             & Tadpole and Symanzik improved \\
Coupling:           & $\beta = 8.0$ \\
Size:               & $20^3 \times 64$ \\
Number$^{a)}$:          & full set: 390 \\
                & reduced set: 150 \\
Lattice spacing$^{b)}$:     & $a = 0.14$\,fm\\
\hline
Smearing parameters$^{c)}$:     & $\alpha_{sm} = 0.55$, $N_{sm} = 25$ \\
HYP parameters$^{d)}$:  & $\alpha_1=1.8$, $\alpha_2=0.6$, $\alpha_3=0.3$ \\
\hline \hline
\end{tabular}
\end{center}
\caption{\label{tab:param} The simulation parameters.
$^{a)}$ The number of uncorrelated configurations. Most data was obtained on the full set of configurations,
and the data for the co-aligned strings was obtained on a reduced set of configurations;
$^{b)}$ Assuming $\sqrt{\sigma}=440$\,MeV;
$^{c)}$ See (\ref{ape_def}) for the definitions;
$^{d)}$ See (\ref{hyp_def}) for the definitions.}
\end{table}

The expectation values for all Wilson loops were measured for the on-axis geometries discussed in the body of the text.

%-----------------------------------------------------------
\subsection{APE Smearing}
\label{app:ape}
We use the traditional 3D APE smearing~\cite{ape}
in order to enhance the overlap of the states, created by the gauge transporters, with the states of the physical
confining strings. It is defined as an iterative fields transformation, one iteration has the form:
\begin{equation}
\label{ape_def}
U^{APE}_{x,i}~=~\left( (1-\alpha_{sm}) U_{x,i} + \frac{\alpha_{sm}}{6}
\sum_{\pm j \neq i}U_{x,j} U_{x+\hat{j},i}
U^\dagger_{x+\hat{i},j} \right)_{SU(3)},
\end{equation}
where $i,j=1 \dots 3$ and $(\cdot)_{SU(3)}$ means a projection onto an $SU(3)$ element. Equation~(\ref{ape_def})
is applied to all spatial links and then $N_{sm}$ iterations are performed. The parameters $\alpha_{sm},~N_{sm}$ are
tuned to compromise CPU time with an enhancement of the ``almost spatial'' Wilson loop $R\times T$ =
$10 \times 1$, see Ref.~\cite{smearparm} for details. Used values are:
\begin{equation}
\alpha_{sm}=0.55~~N_{sm}=25
\end{equation}
The overlap of the state, created by the smeared transporter, and the ground state of the confining
string~\footnote{Do not confuse this ground state of one string with the ground state
of the two strings system, discussed through the paper.}
is consistent with unity ({\it i.e.}, a complete overlap) within the errors for all values of the string
length used.

%-----------------------------------------------------------
\subsection{Hypercubic Blocking}
\label{app:hyp}
The hypercubic blocking (HYP) \cite{hyp1} was proved to be an efficient tool for the static potential
measurements. It is defined as a three step transformation of the time-like links:
%\begin{widetext}
\begin{subequations}
\label{hyp_def}

\begin{eqnarray}
U^{(1)}_{x,\mu;\nu,\rho}&&=~\left( \alpha_1 U_{x,\mu} + \sum_{\pm \eta \neq \mu,\nu,\rho}
U_{x,\eta}U_{x+\hat{\eta},\mu}U_{x+\hat{\mu},\eta}^{\dagger} \right)_{SU(3)} \\
U^{(2)}_{x, \mu;~\nu}&&=~\left( \alpha_2 U_{x, \mu} + \sum_{\pm \rho \neq \mu, \nu}
U^{(1)}_{x,\rho;\nu,\mu} U^{(1)}_{x+\hat{\rho},\mu;\rho,\nu} U^{(1)\dagger}_{x+\hat{\mu},\rho;\nu,\mu}
\right)_{SU(3)} \\
U^{HYP}_{x,\mu}&&=~\left( \alpha_3 U_{x,\mu} + \sum_{\pm \nu \neq \mu}
U^{(2)}_{x,\nu;\mu} U^{(2)}_{x+\hat{\nu},\mu;\nu} U^{(2)\dagger}_{x+\hat{\mu},\nu;\mu}
\right)_{SU(3)},
\end{eqnarray}
\end{subequations}
%\end{widetext}
where $\alpha_{i}$ are parameters of the HYP. We do not iterate the procedure in order to preserve locality.
To tune the HYP  $\alpha_{1,2,3}$ we scan the parameter space for the valley of a maximal enhancement of the ``almost time-like''
Wilson loop $T\times R = 10\times4$. The obtained values of the parameters $\alpha_i$ are:
\begin{equation}
\alpha_1=1.8~~\alpha_2=0.6~~\alpha_3=0.3
\end{equation}
Note that these values are quite different from ``canonical'' ones, obtained by an indirect method,
see Ref.~\cite{hyp1} for details.

The principal effect of the HYP is a suppression of perturbative
contributions to the 2- and 4-quark potentials. For example, the static
potential $V(r)$, calculated after the application of the HYP
procedure with parameters chosen in this paper, has zero quark self-energy
term, a small Coulomb term and a linear term with the unchanged string tension.

The APE and the HYP smearings do not interfere with each other. Therefore the transformations
(\ref{ape_def}) and (\ref{hyp_def}) are both  applied to the initial links.
%\label{app:hyp}

\section{2 and 4-quark potentials}
\label{app:numerical}
\subsection{Remarks on systematical errors}
\label{app:time}
In our paper we use a rather small time extension $t=4a$ of the Wilson loops.
To check the magnitude of the related systematical errors additional
measurements were performed for a larger $t$ with a smaller statistics.

We calculated the 2-quark static potential using the Wilson loops with
time extension $t' = 5 - 8 a$ and the 4-quark potential for the Wilson
loops with $t' = 6a$.

It occurs that the $t$-dependent contribution to $V(r)$ is $0.1$\% from
$t$-independent contribution, see Table \ref{tab:v}. The same ratio for
the 2-string energy levels is of order of $10^{-2}$ (not shown in
tables). Thus we estimate a possible systematic error coming from
finiteness of $t$ to be not higher than $0.1 - 1$\%.

\clearpage

\subsection{Numerical data}

\begin{table*}[!hbpt]
\begin{tabular}{l| r r r r r r r r r r}
\hline \hline
$t/r$ & 1 & 2 & 3 & 4 & 5 & 6 &  7 & 8 & 9 & 10 \\
\hline
& \multicolumn{10}{c}{$aV(r,t=4)$} \\
4 & 0.057987(9) & 0.16662(3) & 0.30149(7) & 0.4209(1) & 0.5331(1) & 0.6420(2) & 0.7493(3) & 0.8560(5) & 0.9625(7) & 1.069(1) \\
\hline
& \multicolumn{10}{c}{$V(r,t) / V(r,t=4)$} \\
5  & 0.9969(2)  & 1.0002(3)  & 1.0016(3)  & 1.0018(4)  & 1.0015(5)  & 1.0010(7)  & 1.0004(8)
    & 1.000(1)  & 1.000(1)  & 1.000(2) \\
6 & 0.9947(2)  & 1.0002(3)  & 1.0026(3)  & 1.0031(5)  & 1.0026(6)  & 1.0017(8)  & 1.001(1)
    & 0.999(1)  & 0.999(2)  & 1.000(3)  \\
7  & 0.9932(2)  & 1.0003(3)  & 1.0034(4)  & 1.0039(5)  & 1.0032(7)  & 1.002(1)  & 1.001(1)
    & 1.001(2)  & 1.003(4)  & 1.009(8)  \\
8  & 0.9920(2)  & 1.0003(3)  & 1.0038(4)  & 1.0042(6)  & 1.0034(9)  & 1.002(1)  & 1.000(2)
    & 1.000(4) & 1.001(8)  & 1.00(1)  \\
\hline \hline
\end{tabular}
\caption{\label{tab:v}Static quark-antiquark potential $V(r)$ (in lattice units) and its
$t$-dependence}
\end{table*}
%
%****************************************************************************************
\begin{table*}[!htbp]
\begin{tabular}{l|r@{.}l r@{.}l r@{.}l r@{.}l r@{.}l r@{.}l r@{.}l r@{.}l r@{.}l}
\hline \hline
$d/r$ &
\multicolumn{2}{c}{2} &
\multicolumn{2}{c}{3} &
\multicolumn{2}{c}{4} &
\multicolumn{2}{c}{5} &
\multicolumn{2}{c}{6} &
\multicolumn{2}{c}{7} &
\multicolumn{2}{c}{8} &
\multicolumn{2}{c}{9} &
\multicolumn{2}{c}{10} \\
\hline
& \multicolumn{18}{c} {Ground state $E_0(r,d)$, GeV} \\
2  &  0&40692(3)  &  0&46497(4)  &  0&46887(3)  &  0&46951(2)  &  0&46963(3)  &
    0&46967(3)  &  0&46968(3)  &  0&46968(3)  &  0&46969(2)\\
3  &  0&46497(4)  &  0&78368(8)  &  0&84369(7)  &  0&84875(5)  &  0&84960(6)  &
    0&84979(6)  &  0&84984(6)  &  0&84984(6)  &  0&84990(4)\\
4  &  0&46887(4)  &  0&84370(8)  &  1&1268(1)  &  1&18117(9)  &  1&1855(1)  &
    1&1862(1)  &  1&1864(1)  &  1&1864(1)  &  1&18657(6)\\
5  &  0&46951(2)  &  0&84875(5)  &  1&18117(9)  &  1&4540(1)  &  1&4992(2)  &
    1&5021(2)  &  1&5027(2)  &  1&5027(2)  &  1&5029(1)\\
6  &  0&46964(3)  &  0&84960(6)  &  1&1854(1)  &  1&4992(2)  &  1&7743(3)  &
    1&8082(3)  &  1&8096(3)  &  1&8097(3)  &  1&8100(2)\\
7  &  0&46967(3)  &  0&84979(6)  &  1&1862(1)  &  1&5021(2)  &  1&8081(3)  &
    2&0910(6)  &  2&1122(5)  &  2&1120(5)  &  2&1129(4)\\
8  &  0&46968(9)  &  0&84984(6)  &  1&1864(1)  &  1&5026(2)  &  1&8096(3)  &
    2&1122(5)  &  2&403(1)  &  2&412(1)  &  2&4142(8)\\
9  &  0&46968(3)  &  0&84984(6)  &  1&1864(1)  &  1&5027(2)  &  1&8097(3)  &
    2&1120(5)  &  2&411(1)  &  2&706(3)  &  2&715(2)\\
10  &  0&46969(2)  &  0&84990(4)  &  1&18658(6)  &  1&5029(1)  &  1&8100(2)  &
    2&1129(4)  &  2&4142(8)  &  2&715(2)  &  3&003(3)\\
\hline
& \multicolumn{18}{c}{ Exited state $E_1(r,d)$, GeV} \\

2  &  0&58197(6)  &  0&9069(1)  &  1&2424(3)  &  1&5563(5)  &  1&852(3)  &  2&127(3)  &
    2&370(6)  &  2&57(1)  &  2&72(1)\\
3  &  0&9068(1)  &  0&9561(1)  &  1&2274(2)  &  1&5332(3)  &  1&829(1)  &  2&111(2)  &  2&376(4)
    &  2&616(9)  &  2&83(1)\\
4  &  1&2421(3)  &  1&2273(2)  &  1&2711(2)  &  1&5264(2)  &  1&8216(7)  &  2&113(1)  &  2&397(3)
    &  2&671(7)  &  2&932(9)\\
5  &  1&5558(5)  &  1&5330(3)  &  1&5264(2)  &  1&5638(2)  &  1&8205(4)  &  2&1147(9)  &  2&408(2)
    &  2&699(5)  &  2&982(7)\\
6  &  1&851(2)  &  1&829(1)  &  1&8218(7)  &  1&8207(4)  &  1&8500(4)  &  2&1165(7)  &  2&413(2)
    &  2&710(4)  &  3&004(6)\\
7  &  2&124(3)  &  2&111(2)  &  2&113(1)  &  2&1153(9)  &  2&1168(7)  &  2&1367(6)  &  2&416(1)  &
    2&715(3)  &  3&017(5)\\
8  &  2&366(6)  &  2&375(5)  &  2&397(3)  &  2&408(2)  &  2&414(2)  &  2&416(1)  &  2&424(1)  &
    2&713(3)  &  3&014(5)\\
9  &  2&56(1)  &  2&613(9)  &  2&671(7)  &  2&699(5)  &  2&709(4)  &  2&715(3)  &  2&712(3)  &
    2&715(3)  &  3&010(5)\\
10  &  2&71(1)  &  2&82(1)  &  2&929(9)  &  2&982(7)  &  3&005(6)  &  3&017(5)  &  3&014(5)  &
    3&009(5)  &  3&030(4)\\
\hline \hline
\end{tabular}
\caption{\label{tab:e.opp}Energy levels of oppositely aligned strings as functions of geometry $r,d$
(in lattice units, $a = 0.14$\,fm).}
\end{table*}
%
%---------------------------------------------------------------------------
\begin{table*}[!htbp]
\begin{tabular}{l | r@{.}l r@{.}l r@{.}l r@{.}l @{~~}| @{~~}r@{.}l r@{.}l r@{.}l r@{.}l}
\hline \hline
$d/r$ &
\multicolumn{2}{c}{4} &
\multicolumn{2}{c}{6} &
\multicolumn{2}{c}{8} &
\multicolumn{2}{c|@{~~}}{10} &
\multicolumn{2}{c}{4} &
\multicolumn{2}{c}{6} &
\multicolumn{2}{c}{8} &
\multicolumn{2}{c}{10} \\
\hline
& \multicolumn{8}{c|@{~~}}{Ground state $E_0(r,d)$, GeV} &
\multicolumn{8}{c}{Exited state $E_1(r,d)$, GeV} \\

2  &  0&8829(3)  &  1&1943(4)  &  1&4950(7)  &  1&795(1) &
     1&347(1)  &  1&983(4)  &  2&58(1)  &  3&12(5)\\
3  &  1&0909(3)  &  1&4450(5)  &  1&7509(7)  &  2&052(1) &
    1&3689(7)  &  1&926(2)  &  2&495(6)  &  3&03(2)\\
4  &  1&1683(3)  &  1&6571(5)  &  1&9856(9)  &  2&291(2) &
    1&5284(8)  &  1&923(1)  &  2&470(3)  &  3&03(1)\\
5  &  1&1828(2)  &  1&7796(5)  &  2&226(1)  &  2&549(2) &
    1&764(1)  &  2&038(1)  &  2&469(2)  &  3&028(9)\\
6  &  1&1855(2)  &  1&8036(4)  &  2&382(2)  &  2&800(5) &
    2&002(2)  &  2&251(2)  &  2&554(3)  &  3&052(9)\\
7  &  1&1862(1)  &  1&8085(4)  &  2&411(2)  &  2&995(7) &
    2&234(3)  &  2&500(3)  &  2&789(4)  &  3&14(1)\\
8  &  1&1864(1)  &  1&8097(4)  &  2&417(1)  &  3&035(8) &
    2&445(4)  &  2&746(5)  &  3&047(9)  &  3&36(2)\\
9  &  1&1864(1)  &  1&8100(3)  &  2&421(1)  &  3&042(8) &
    2&641(7)  &  3&00(1)  &  3&32(2)  &  3&60(4)\\
10  &  1&18655(8)  &  1&8097(3)  &  2&416(1)  &  3&033(7) &
    2&814(8)  &  3&24(2)  &  3&57(3)  &  3&82(7)\\
\hline \hline
\end{tabular}
\caption{\label{tab:e.same} The same as in Table~\ref{tab:e.opp} but for the co-aligned strings.}
\end{table*}

\clearpage

%===========================================================

\end{document}

%% file: eps/energy.opp.tex
%GNUPLOT: LaTeX picture with Postscript
\begin{picture}(0,0)%
\includegraphics{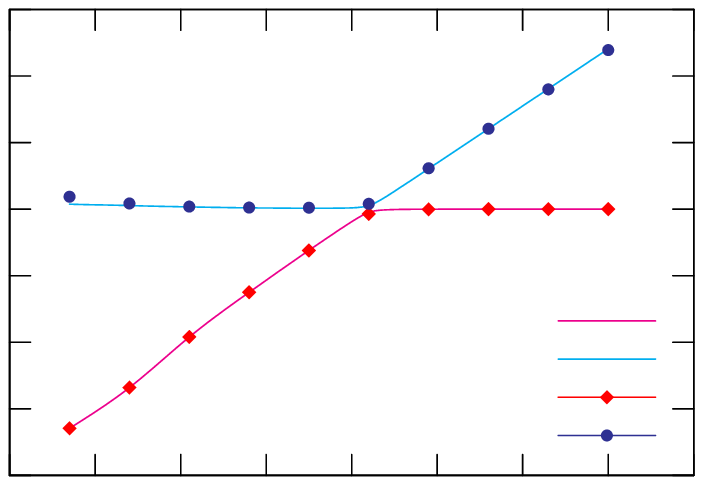}%
\end{picture}%
\begingroup
\setlength{\unitlength}{0.0200bp}%
\begin{picture}(12599,8909)(0,0)%
\put(1650,1650){\makebox(0,0)[r]{\strut{}-2}}%
\put(1650,2609){\makebox(0,0)[r]{\strut{}-1.5}}%
\put(1650,3567){\makebox(0,0)[r]{\strut{}-1}}%
\put(1650,4526){\makebox(0,0)[r]{\strut{}-0.5}}%
\put(1650,5484){\makebox(0,0)[r]{\strut{} 0}}%
\put(1650,6443){\makebox(0,0)[r]{\strut{} 0.5}}%
\put(1650,7401){\makebox(0,0)[r]{\strut{} 1}}%
\put(1650,8360){\makebox(0,0)[r]{\strut{} 1.5}}%
\put(1925,1100){\makebox(0,0){\strut{} 0}}%
\put(3156,1100){\makebox(0,0){\strut{} 0.2}}%
\put(4388,1100){\makebox(0,0){\strut{} 0.4}}%
\put(5619,1100){\makebox(0,0){\strut{} 0.6}}%
\put(6850,1100){\makebox(0,0){\strut{} 0.8}}%
\put(8081,1100){\makebox(0,0){\strut{} 1}}%
\put(9312,1100){\makebox(0,0){\strut{} 1.2}}%
\put(10544,1100){\makebox(0,0){\strut{} 1.4}}%
\put(11775,1100){\makebox(0,0){\strut{} 1.6}}%
\put(550,5005){\rotatebox{90}{\makebox(0,0){\strut{}$E-2V(r)$,\,GeV}}}%
\put(6850,275){\makebox(0,0){\strut{}$d$,\,fm}}%
\put(9550,3875){\makebox(0,0)[r]{\strut{}$E_0$, soapfilm}}%
\put(9550,3325){\makebox(0,0)[r]{\strut{}$E_1$, soapfilm}}%
\put(9550,2775){\makebox(0,0)[r]{\strut{}$E_0$, data}}%
\put(9550,2225){\makebox(0,0)[r]{\strut{}$E_1$, data}}%
\end{picture}%
\endgroup
 

%% file: eps/energy.same.tex
%GNUPLOT: LaTeX picture with Postscript
\begin{picture}(0,0)%
\includegraphics{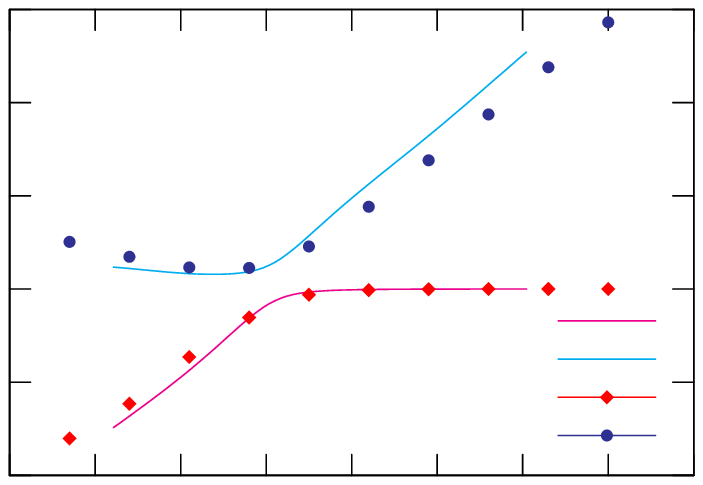}%
\end{picture}%
\begingroup
\setlength{\unitlength}{0.0200bp}%
\begin{picture}(12599,8909)(0,0)%
\put(1650,1650){\makebox(0,0)[r]{\strut{}-1}}%
\put(1650,2992){\makebox(0,0)[r]{\strut{}-0.5}}%
\put(1650,4334){\makebox(0,0)[r]{\strut{} 0}}%
\put(1650,5676){\makebox(0,0)[r]{\strut{} 0.5}}%
\put(1650,7018){\makebox(0,0)[r]{\strut{} 1}}%
\put(1650,8360){\makebox(0,0)[r]{\strut{} 1.5}}%
\put(1925,1100){\makebox(0,0){\strut{} 0}}%
\put(3156,1100){\makebox(0,0){\strut{} 0.2}}%
\put(4388,1100){\makebox(0,0){\strut{} 0.4}}%
\put(5619,1100){\makebox(0,0){\strut{} 0.6}}%
\put(6850,1100){\makebox(0,0){\strut{} 0.8}}%
\put(8081,1100){\makebox(0,0){\strut{} 1}}%
\put(9312,1100){\makebox(0,0){\strut{} 1.2}}%
\put(10544,1100){\makebox(0,0){\strut{} 1.4}}%
\put(11775,1100){\makebox(0,0){\strut{} 1.6}}%
\put(550,5005){\rotatebox{90}{\makebox(0,0){\strut{}$E-2V(r)$,\,GeV}}}%
\put(6850,275){\makebox(0,0){\strut{}$d$,\,fm}}%
\put(9550,3875){\makebox(0,0)[r]{\strut{}$E_0$, soapfilm}}%
\put(9550,3325){\makebox(0,0)[r]{\strut{}$E_1$, soapfilm}}%
\put(9550,2775){\makebox(0,0)[r]{\strut{}$E_0$, data}}%
\put(9550,2225){\makebox(0,0)[r]{\strut{}$E_1$, data}}%
\end{picture}%
\endgroup
 

%% file: eps/ed.0.tex
%GNUPLOT: LaTeX picture with Postscript
\begin{picture}(0,0)%
\includegraphics{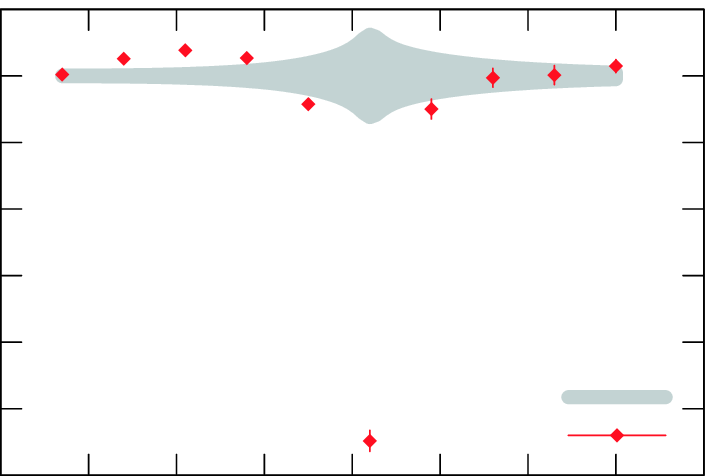}%
\end{picture}%
\begingroup
\setlength{\unitlength}{0.0200bp}%
\begin{picture}(12599,8909)(0,0)%
\put(1375,1650){\makebox(0,0)[r]{\strut{}-12}}%
\put(1375,2609){\makebox(0,0)[r]{\strut{}-10}}%
\put(1375,3567){\makebox(0,0)[r]{\strut{}-8}}%
\put(1375,4526){\makebox(0,0)[r]{\strut{}-6}}%
\put(1375,5484){\makebox(0,0)[r]{\strut{}-4}}%
\put(1375,6443){\makebox(0,0)[r]{\strut{}-2}}%
\put(1375,7401){\makebox(0,0)[r]{\strut{} 0}}%
\put(1375,8360){\makebox(0,0)[r]{\strut{} 2}}%
\put(1650,1100){\makebox(0,0){\strut{} 0}}%
\put(2916,1100){\makebox(0,0){\strut{} 0.2}}%
\put(4181,1100){\makebox(0,0){\strut{} 0.4}}%
\put(5447,1100){\makebox(0,0){\strut{} 0.6}}%
\put(6713,1100){\makebox(0,0){\strut{} 0.8}}%
\put(7978,1100){\makebox(0,0){\strut{} 1}}%
\put(9244,1100){\makebox(0,0){\strut{} 1.2}}%
\put(10509,1100){\makebox(0,0){\strut{} 1.4}}%
\put(11775,1100){\makebox(0,0){\strut{} 1.6}}%
\put(550,5005){\rotatebox{90}{\makebox(0,0){\strut{}$E_{data}-E_{SF}$,\,MeV}}}%
\put(6712,275){\makebox(0,0){\strut{}$d$,\,fm}}%
\put(9550,2775){\makebox(0,0)[r]{\strut{}$\sigma_{SF}$}}%
\put(9550,2225){\makebox(0,0)[r]{\strut{}$\delta E_0$}}%
\end{picture}%
\endgroup
 

%% file: eps/ed.1.tex
%GNUPLOT: LaTeX picture with Postscript
\begin{picture}(0,0)%
\includegraphics{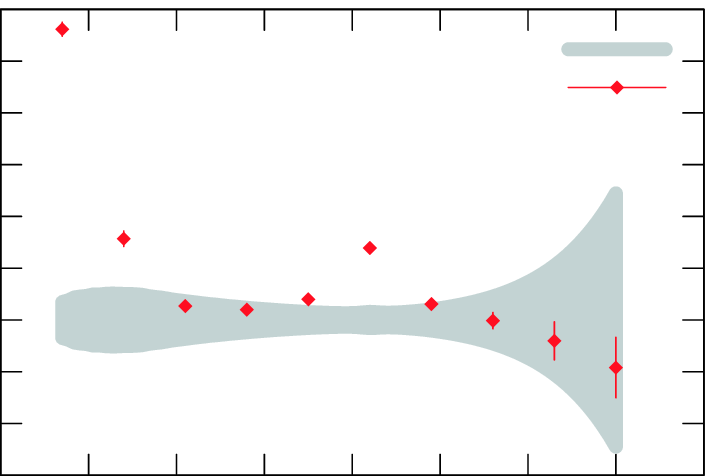}%
\end{picture}%
\begingroup
\setlength{\unitlength}{0.0200bp}%
\begin{picture}(12599,8909)(0,0)%
\put(1375,1650){\makebox(0,0)[r]{\strut{}-30}}%
\put(1375,2396){\makebox(0,0)[r]{\strut{}-20}}%
\put(1375,3141){\makebox(0,0)[r]{\strut{}-10}}%
\put(1375,3887){\makebox(0,0)[r]{\strut{} 0}}%
\put(1375,4632){\makebox(0,0)[r]{\strut{} 10}}%
\put(1375,5378){\makebox(0,0)[r]{\strut{} 20}}%
\put(1375,6123){\makebox(0,0)[r]{\strut{} 30}}%
\put(1375,6869){\makebox(0,0)[r]{\strut{} 40}}%
\put(1375,7614){\makebox(0,0)[r]{\strut{} 50}}%
\put(1375,8360){\makebox(0,0)[r]{\strut{} 60}}%
\put(1650,1100){\makebox(0,0){\strut{} 0}}%
\put(2916,1100){\makebox(0,0){\strut{} 0.2}}%
\put(4181,1100){\makebox(0,0){\strut{} 0.4}}%
\put(5447,1100){\makebox(0,0){\strut{} 0.6}}%
\put(6713,1100){\makebox(0,0){\strut{} 0.8}}%
\put(7978,1100){\makebox(0,0){\strut{} 1}}%
\put(9244,1100){\makebox(0,0){\strut{} 1.2}}%
\put(10509,1100){\makebox(0,0){\strut{} 1.4}}%
\put(11775,1100){\makebox(0,0){\strut{} 1.6}}%
\put(550,5005){\rotatebox{90}{\makebox(0,0){\strut{}$E_{data}-E_{SF}$,\,MeV}}}%
\put(6712,275){\makebox(0,0){\strut{}$d$,\,fm}}%
\put(9550,7785){\makebox(0,0)[r]{\strut{}$\sigma_{SF}$}}%
\put(9550,7235){\makebox(0,0)[r]{\strut{}$\delta E_1$}}%
\end{picture}%
\endgroup
 

%% file: eps/ed.3.tex
%GNUPLOT: LaTeX picture with Postscript
\begin{picture}(0,0)%
\includegraphics{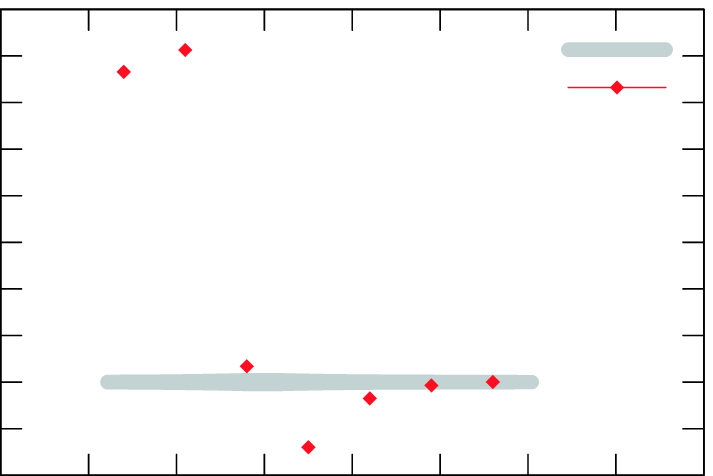}%
\end{picture}%
\begingroup
\setlength{\unitlength}{0.0200bp}%
\begin{picture}(12599,8909)(0,0)%
\put(1375,1650){\makebox(0,0)[r]{\strut{}-20}}%
\put(1375,2321){\makebox(0,0)[r]{\strut{}-10}}%
\put(1375,2992){\makebox(0,0)[r]{\strut{} 0}}%
\put(1375,3663){\makebox(0,0)[r]{\strut{} 10}}%
\put(1375,4334){\makebox(0,0)[r]{\strut{} 20}}%
\put(1375,5005){\makebox(0,0)[r]{\strut{} 30}}%
\put(1375,5676){\makebox(0,0)[r]{\strut{} 40}}%
\put(1375,6347){\makebox(0,0)[r]{\strut{} 50}}%
\put(1375,7018){\makebox(0,0)[r]{\strut{} 60}}%
\put(1375,7689){\makebox(0,0)[r]{\strut{} 70}}%
\put(1375,8360){\makebox(0,0)[r]{\strut{} 80}}%
\put(1650,1100){\makebox(0,0){\strut{} 0}}%
\put(2916,1100){\makebox(0,0){\strut{} 0.2}}%
\put(4181,1100){\makebox(0,0){\strut{} 0.4}}%
\put(5447,1100){\makebox(0,0){\strut{} 0.6}}%
\put(6713,1100){\makebox(0,0){\strut{} 0.8}}%
\put(7978,1100){\makebox(0,0){\strut{} 1}}%
\put(9244,1100){\makebox(0,0){\strut{} 1.2}}%
\put(10509,1100){\makebox(0,0){\strut{} 1.4}}%
\put(11775,1100){\makebox(0,0){\strut{} 1.6}}%
\put(550,5005){\rotatebox{90}{\makebox(0,0){\strut{}$E_{data}-E_{SF}$,\,MeV}}}%
\put(6712,275){\makebox(0,0){\strut{}$d$,\,fm}}%
\put(9550,7785){\makebox(0,0)[r]{\strut{}$\sigma_{SF}$}}%
\put(9550,7235){\makebox(0,0)[r]{\strut{}$\delta E_0$}}%
\end{picture}%
\endgroup
 

%% file: eps/ed.4.tex
%GNUPLOT: LaTeX picture with Postscript
\begin{picture}(0,0)%
\includegraphics{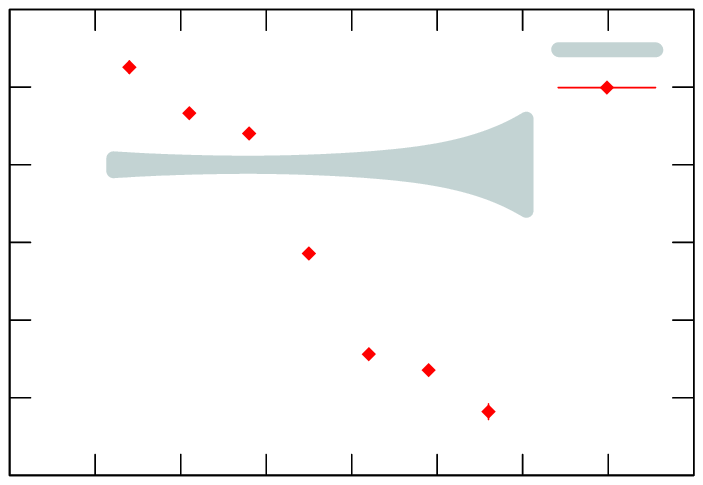}%
\end{picture}%
\begingroup
\setlength{\unitlength}{0.0200bp}%
\begin{picture}(12599,8909)(0,0)%
\put(1650,1650){\makebox(0,0)[r]{\strut{}-200}}%
\put(1650,2768){\makebox(0,0)[r]{\strut{}-150}}%
\put(1650,3887){\makebox(0,0)[r]{\strut{}-100}}%
\put(1650,5005){\makebox(0,0)[r]{\strut{}-50}}%
\put(1650,6123){\makebox(0,0)[r]{\strut{} 0}}%
\put(1650,7242){\makebox(0,0)[r]{\strut{} 50}}%
\put(1650,8360){\makebox(0,0)[r]{\strut{} 100}}%
\put(1925,1100){\makebox(0,0){\strut{} 0}}%
\put(3156,1100){\makebox(0,0){\strut{} 0.2}}%
\put(4388,1100){\makebox(0,0){\strut{} 0.4}}%
\put(5619,1100){\makebox(0,0){\strut{} 0.6}}%
\put(6850,1100){\makebox(0,0){\strut{} 0.8}}%
\put(8081,1100){\makebox(0,0){\strut{} 1}}%
\put(9312,1100){\makebox(0,0){\strut{} 1.2}}%
\put(10544,1100){\makebox(0,0){\strut{} 1.4}}%
\put(11775,1100){\makebox(0,0){\strut{} 1.6}}%
\put(550,5005){\rotatebox{90}{\makebox(0,0){\strut{}$E_{data}-E_{SF}$,\,MeV}}}%
\put(6850,275){\makebox(0,0){\strut{}$d$,\,fm}}%
\put(9550,7785){\makebox(0,0)[r]{\strut{}$\sigma_{SF}$}}%
\put(9550,7235){\makebox(0,0)[r]{\strut{}$\delta E_1$}}%
\end{picture}%
\endgroup
 

%% file: eps/split.t4.tex
%GNUPLOT: LaTeX picture with Postscript
\begin{picture}(0,0)%
\includegraphics{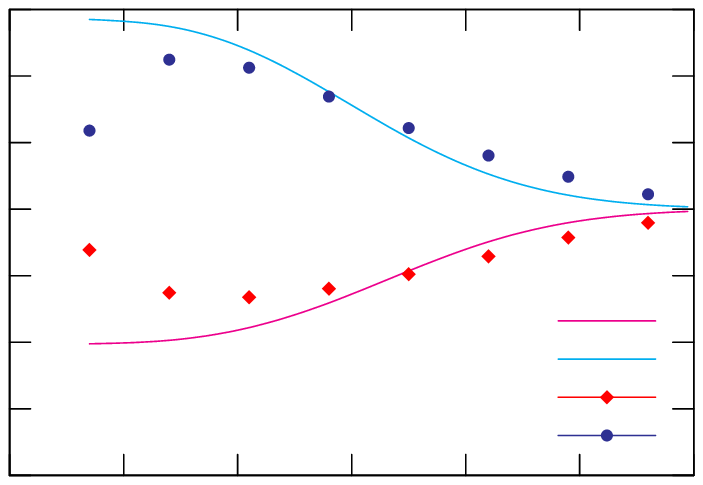}%
\end{picture}%
\begingroup
\setlength{\unitlength}{0.0200bp}%
\begin{picture}(12599,8909)(0,0)%
\put(1650,1650){\makebox(0,0)[r]{\strut{}-200}}%
\put(1650,2609){\makebox(0,0)[r]{\strut{}-150}}%
\put(1650,3567){\makebox(0,0)[r]{\strut{}-100}}%
\put(1650,4526){\makebox(0,0)[r]{\strut{}-50}}%
\put(1650,5484){\makebox(0,0)[r]{\strut{} 0}}%
\put(1650,6443){\makebox(0,0)[r]{\strut{} 50}}%
\put(1650,7401){\makebox(0,0)[r]{\strut{} 100}}%
\put(1650,8360){\makebox(0,0)[r]{\strut{} 150}}%
\put(1925,1100){\makebox(0,0){\strut{} 0}}%
\put(3567,1100){\makebox(0,0){\strut{} 0.2}}%
\put(5208,1100){\makebox(0,0){\strut{} 0.4}}%
\put(6850,1100){\makebox(0,0){\strut{} 0.6}}%
\put(8492,1100){\makebox(0,0){\strut{} 0.8}}%
\put(10133,1100){\makebox(0,0){\strut{} 1}}%
\put(11775,1100){\makebox(0,0){\strut{} 1.2}}%
\put(550,5005){\rotatebox{90}{\makebox(0,0){\strut{}$E(r,r) - 2V(r)$,\,MeV}}}%
\put(6850,275){\makebox(0,0){\strut{}$r$\,,fm}}%
\put(9550,3875){\makebox(0,0)[r]{\strut{}ground, SF}}%
\put(9550,3325){\makebox(0,0)[r]{\strut{}exited, SF}}%
\put(9550,2775){\makebox(0,0)[r]{\strut{}ground, data}}%
\put(9550,2225){\makebox(0,0)[r]{\strut{}exited, data}}%
\end{picture}%
\endgroup
 

%% file: eps/split.t6.tex
%GNUPLOT: LaTeX picture with Postscript
\begin{picture}(0,0)%
\includegraphics{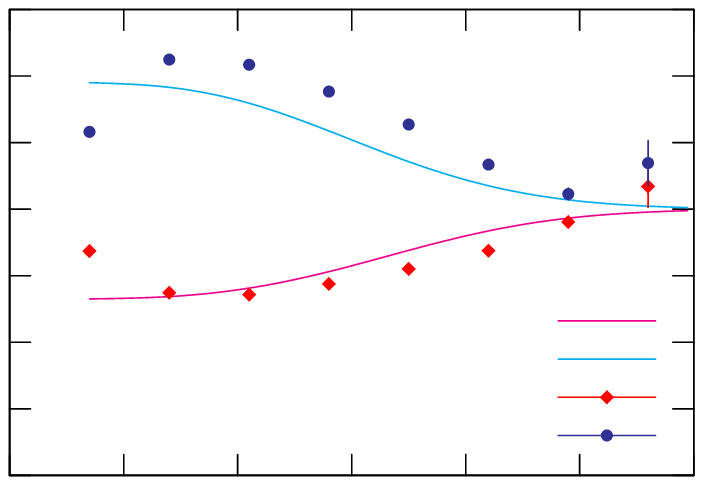}%
\end{picture}%
\begingroup
\setlength{\unitlength}{0.0200bp}%
\begin{picture}(12599,8909)(0,0)%
\put(1650,1650){\makebox(0,0)[r]{\strut{}-200}}%
\put(1650,2609){\makebox(0,0)[r]{\strut{}-150}}%
\put(1650,3567){\makebox(0,0)[r]{\strut{}-100}}%
\put(1650,4526){\makebox(0,0)[r]{\strut{}-50}}%
\put(1650,5484){\makebox(0,0)[r]{\strut{} 0}}%
\put(1650,6443){\makebox(0,0)[r]{\strut{} 50}}%
\put(1650,7401){\makebox(0,0)[r]{\strut{} 100}}%
\put(1650,8360){\makebox(0,0)[r]{\strut{} 150}}%
\put(1925,1100){\makebox(0,0){\strut{} 0}}%
\put(3567,1100){\makebox(0,0){\strut{} 0.2}}%
\put(5208,1100){\makebox(0,0){\strut{} 0.4}}%
\put(6850,1100){\makebox(0,0){\strut{} 0.6}}%
\put(8492,1100){\makebox(0,0){\strut{} 0.8}}%
\put(10133,1100){\makebox(0,0){\strut{} 1}}%
\put(11775,1100){\makebox(0,0){\strut{} 1.2}}%
\put(550,5005){\rotatebox{90}{\makebox(0,0){\strut{}$E(r,r) - 2V(r)$,\,MeV}}}%
\put(6850,275){\makebox(0,0){\strut{}$r$\,,fm}}%
\put(9550,3875){\makebox(0,0)[r]{\strut{}ground, SF}}%
\put(9550,3325){\makebox(0,0)[r]{\strut{}exited, SF}}%
\put(9550,2775){\makebox(0,0)[r]{\strut{}ground, data}}%
\put(9550,2225){\makebox(0,0)[r]{\strut{}exited, data}}%
\end{picture}%
\endgroup
 

%% file: eps/h0.opp.tex
%GNUPLOT: LaTeX picture with Postscript
\begin{picture}(0,0)%
\includegraphics{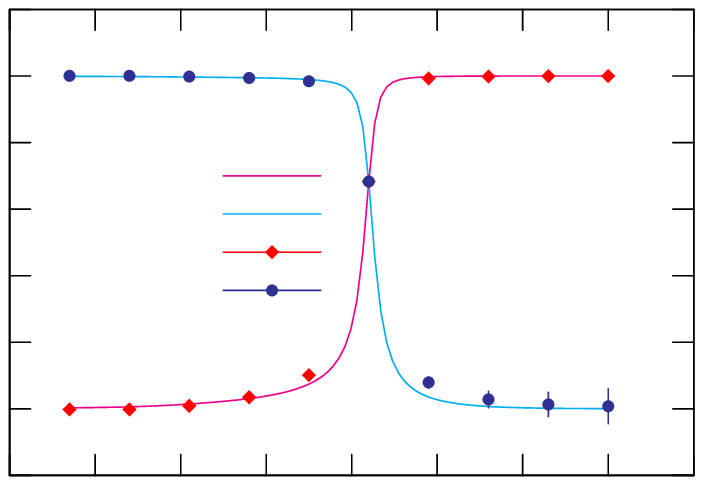}%
\end{picture}%
\begingroup
\setlength{\unitlength}{0.0200bp}%
\begin{picture}(12599,8909)(0,0)%
\put(1650,1650){\makebox(0,0)[r]{\strut{}-0.2}}%
\put(1650,2609){\makebox(0,0)[r]{\strut{} 0}}%
\put(1650,3567){\makebox(0,0)[r]{\strut{} 0.2}}%
\put(1650,4526){\makebox(0,0)[r]{\strut{} 0.4}}%
\put(1650,5484){\makebox(0,0)[r]{\strut{} 0.6}}%
\put(1650,6443){\makebox(0,0)[r]{\strut{} 0.8}}%
\put(1650,7401){\makebox(0,0)[r]{\strut{} 1}}%
\put(1650,8360){\makebox(0,0)[r]{\strut{} 1.2}}%
\put(1925,1100){\makebox(0,0){\strut{} 0}}%
\put(3156,1100){\makebox(0,0){\strut{} 0.2}}%
\put(4388,1100){\makebox(0,0){\strut{} 0.4}}%
\put(5619,1100){\makebox(0,0){\strut{} 0.6}}%
\put(6850,1100){\makebox(0,0){\strut{} 0.8}}%
\put(8081,1100){\makebox(0,0){\strut{} 1}}%
\put(9312,1100){\makebox(0,0){\strut{} 1.2}}%
\put(10544,1100){\makebox(0,0){\strut{} 1.4}}%
\put(11775,1100){\makebox(0,0){\strut{} 1.6}}%
\put(550,5005){\rotatebox{90}{\makebox(0,0){\strut{}$h_{0i}$}}}%
\put(6850,275){\makebox(0,0){\strut{}$d$,\,fm}}%
\put(4728,5964){\makebox(0,0)[r]{\strut{}$h_{0A}$, SF}}%
\put(4728,5414){\makebox(0,0)[r]{\strut{}$h_{0B}$, SF}}%
\put(4728,4864){\makebox(0,0)[r]{\strut{}$h_{0A}$, data}}%
\put(4728,4314){\makebox(0,0)[r]{\strut{}$h_{0B}$, data}}%
\end{picture}%
\endgroup
 

%% file: eps/h0.same.tex
%GNUPLOT: LaTeX picture with Postscript
\begin{picture}(0,0)%
\includegraphics{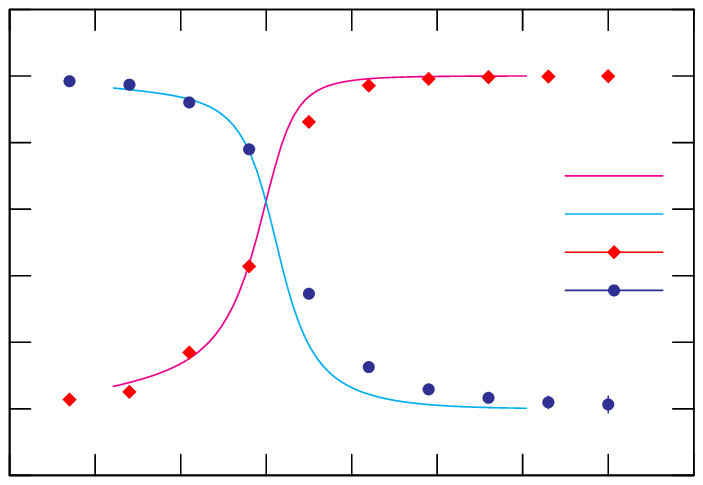}%
\end{picture}%
\begingroup
\setlength{\unitlength}{0.0200bp}%
\begin{picture}(12599,8909)(0,0)%
\put(1650,1650){\makebox(0,0)[r]{\strut{}-0.2}}%
\put(1650,2609){\makebox(0,0)[r]{\strut{} 0}}%
\put(1650,3567){\makebox(0,0)[r]{\strut{} 0.2}}%
\put(1650,4526){\makebox(0,0)[r]{\strut{} 0.4}}%
\put(1650,5484){\makebox(0,0)[r]{\strut{} 0.6}}%
\put(1650,6443){\makebox(0,0)[r]{\strut{} 0.8}}%
\put(1650,7401){\makebox(0,0)[r]{\strut{} 1}}%
\put(1650,8360){\makebox(0,0)[r]{\strut{} 1.2}}%
\put(1925,1100){\makebox(0,0){\strut{} 0}}%
\put(3156,1100){\makebox(0,0){\strut{} 0.2}}%
\put(4388,1100){\makebox(0,0){\strut{} 0.4}}%
\put(5619,1100){\makebox(0,0){\strut{} 0.6}}%
\put(6850,1100){\makebox(0,0){\strut{} 0.8}}%
\put(8081,1100){\makebox(0,0){\strut{} 1}}%
\put(9312,1100){\makebox(0,0){\strut{} 1.2}}%
\put(10544,1100){\makebox(0,0){\strut{} 1.4}}%
\put(11775,1100){\makebox(0,0){\strut{} 1.6}}%
\put(550,5005){\rotatebox{90}{\makebox(0,0){\strut{}$h_{0i}$}}}%
\put(6850,275){\makebox(0,0){\strut{}$d$,\,fm}}%
\put(9653,5964){\makebox(0,0)[r]{\strut{}$h_{0A}$, SF}}%
\put(9653,5414){\makebox(0,0)[r]{\strut{}$h_{0B}$, SF}}%
\put(9653,4864){\makebox(0,0)[r]{\strut{}$h_{0A}$, data}}%
\put(9653,4314){\makebox(0,0)[r]{\strut{}$h_{0B}$, data}}%
\end{picture}%
\endgroup
 

%% file: eps/h1.opp.tex
%GNUPLOT: LaTeX picture with Postscript
\begin{picture}(0,0)%
\includegraphics{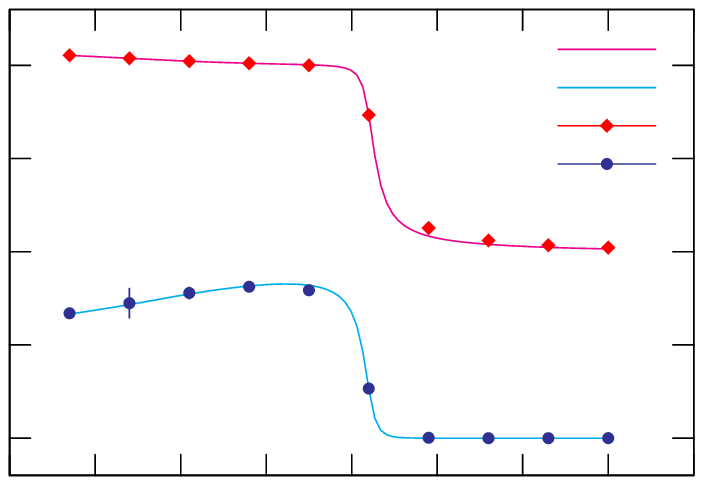}%
\end{picture}%
\begingroup
\setlength{\unitlength}{0.0200bp}%
\begin{picture}(12599,8909)(0,0)%
\put(1650,2187){\makebox(0,0)[r]{\strut{}-1}}%
\put(1650,3529){\makebox(0,0)[r]{\strut{}-0.5}}%
\put(1650,4871){\makebox(0,0)[r]{\strut{} 0}}%
\put(1650,6213){\makebox(0,0)[r]{\strut{} 0.5}}%
\put(1650,7555){\makebox(0,0)[r]{\strut{} 1}}%
\put(1925,1100){\makebox(0,0){\strut{} 0}}%
\put(3156,1100){\makebox(0,0){\strut{} 0.2}}%
\put(4388,1100){\makebox(0,0){\strut{} 0.4}}%
\put(5619,1100){\makebox(0,0){\strut{} 0.6}}%
\put(6850,1100){\makebox(0,0){\strut{} 0.8}}%
\put(8081,1100){\makebox(0,0){\strut{} 1}}%
\put(9312,1100){\makebox(0,0){\strut{} 1.2}}%
\put(10544,1100){\makebox(0,0){\strut{} 1.4}}%
\put(11775,1100){\makebox(0,0){\strut{} 1.6}}%
\put(550,5005){\rotatebox{90}{\makebox(0,0){\strut{}$h_{1i}$}}}%
\put(6850,275){\makebox(0,0){\strut{}$d$,\,fm}}%
\put(9550,7785){\makebox(0,0)[r]{\strut{}$h_{1A}$, SF}}%
\put(9550,7235){\makebox(0,0)[r]{\strut{}$h_{1B}$, SF}}%
\put(9550,6685){\makebox(0,0)[r]{\strut{}$h_{1A}$, data}}%
\put(9550,6135){\makebox(0,0)[r]{\strut{}$h_{1B}$, data}}%
\end{picture}%
\endgroup
 

%% file: eps/h1.same.tex
%GNUPLOT: LaTeX picture with Postscript
\begin{picture}(0,0)%
\includegraphics{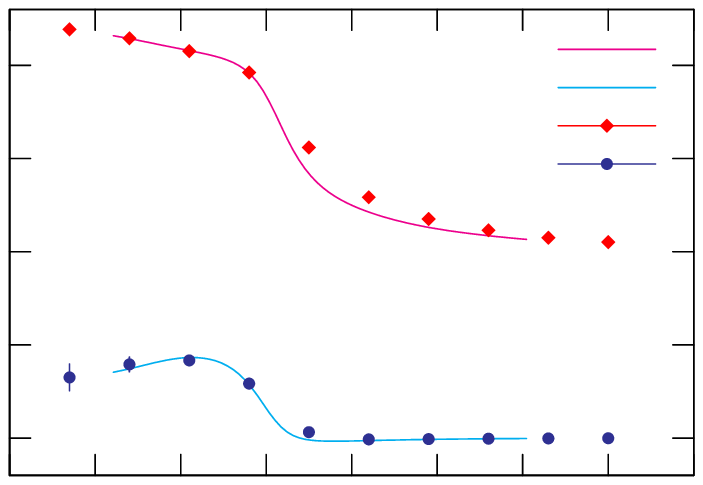}%
\end{picture}%
\begingroup
\setlength{\unitlength}{0.0200bp}%
\begin{picture}(12599,8909)(0,0)%
\put(1650,2187){\makebox(0,0)[r]{\strut{}-1}}%
\put(1650,3529){\makebox(0,0)[r]{\strut{}-0.5}}%
\put(1650,4871){\makebox(0,0)[r]{\strut{} 0}}%
\put(1650,6213){\makebox(0,0)[r]{\strut{} 0.5}}%
\put(1650,7555){\makebox(0,0)[r]{\strut{} 1}}%
\put(1925,1100){\makebox(0,0){\strut{} 0}}%
\put(3156,1100){\makebox(0,0){\strut{} 0.2}}%
\put(4388,1100){\makebox(0,0){\strut{} 0.4}}%
\put(5619,1100){\makebox(0,0){\strut{} 0.6}}%
\put(6850,1100){\makebox(0,0){\strut{} 0.8}}%
\put(8081,1100){\makebox(0,0){\strut{} 1}}%
\put(9312,1100){\makebox(0,0){\strut{} 1.2}}%
\put(10544,1100){\makebox(0,0){\strut{} 1.4}}%
\put(11775,1100){\makebox(0,0){\strut{} 1.6}}%
\put(550,5005){\rotatebox{90}{\makebox(0,0){\strut{}$h_{1i}$}}}%
\put(6850,275){\makebox(0,0){\strut{}$d$,\,fm}}%
\put(9550,7785){\makebox(0,0)[r]{\strut{}$h_{1A}$, SF}}%
\put(9550,7235){\makebox(0,0)[r]{\strut{}$h_{1B}$, SF}}%
\put(9550,6685){\makebox(0,0)[r]{\strut{}$h_{1A}$, data}}%
\put(9550,6135){\makebox(0,0)[r]{\strut{}$h_{1B}$, data}}%
\end{picture}%
\endgroup
 

%% file: str2.bbl
\begin{thebibliography}{99}

\bibitem{ref:Reviews:QCDforces}
W.~Lucha, F.~F.~Schoberl and D.~Gromes,
%``Bound states of quarks,''
Phys.\ Rept.\  {\bf 200} (1991) 127;\,
%%CITATION = PRPLC,200,127;%%
  G.~S.~Bali,
  %``QCD forces and heavy quark bound states,''
  Phys.\ Rept.\  {\bf 343}, 1 (2001).
%%CITATION = HEP-PH 0001312;%%

\bibitem{ref:Reviews:correlators}
  A.~Di Giacomo, H.~G.~Dosch, V.~I.~Shevchenko and Yu.~A.~Simonov,
  %``Field correlators in QCD: Theory and applications,''
  Phys.\ Rept.\  {\bf 372}, 319 (2002).
%%CITATION = HEP-PH 0007223;%%

\bibitem{Takahashi:2002bw}
  T.~T.~Takahashi, H.~Matsufuru, Y.~Nemoto and H.~Suganuma,
  %``The three-quark potential in the SU(3) lattice QCD,''
  Phys.\ Rev.\ Lett.\  {\bf 86}, 18 (2001);
%%CITATION = HEP-LAT 0006005;%%
  T.~T.~Takahashi, H.~Suganuma, Y.~Nemoto and H.~Matsufuru,
  %``Detailed analysis of the three quark potential in SU(3) lattice QCD,''
  Phys.\ Rev.\ D {\bf 65}, 114509 (2002).
%%CITATION = HEP-LAT 0204011;%%

\bibitem{Alexandrou:2002sn}
C.~Alexandrou, P.~De Forcrand and A.~Tsapalis,
  %``The static three-quark SU(3) and four-quark SU(4) potentials,''
  Phys.\ Rev.\ D {\bf 65}, 054503 (2002).
%%CITATION = HEP-LAT 0107006;%%

\bibitem{Bornyakov:2004uv}
V.~G.~Bornyakov {\it et al.}  [DIK Collaboration],
%``Baryonic flux in quenched and two-flavor dynamical QCD,''
Phys.\ Rev.\ D {\bf 70}, 054506 (2004).
%%CITATION = HEP-LAT 0401026;%%

\bibitem{ref:Ydominance}
F.~Okiharu, H.~Suganuma and T.~T.~Takahashi,
  %``The tetraquark potential and flip-flop in SU(3) lattice QCD,''
  Phys.\ Rev.\ D {\bf 72}, 014505 (2005).
%%CITATION = HEP-LAT 0412012;%%

\bibitem{ref:profile:DIK}
H.~Ichie, V.~Bornyakov, T.~Streuer and G.~Schierholz,
  %``Flux tubes of two- and three-quark system in full QCD,''
  Nucl.\ Phys.\ A {\bf 721}, 899 (2003);
%%CITATION = HEP-LAT 0212036;%%
  V.~G.~Bornyakov {\it et al.},
  %``Profiles of the broken string in two-flavor QCD below and above the  finite
  %temperature transition,''
  Prog.\ Theor.\ Phys.\  {\bf 112}, 307 (2004).
%%CITATION = HEP-LAT 0401027;%%

\bibitem{Bissey:2005sk}
F.~Bissey {\it et al.},
%``Gluon field distribution in baryons,''
Nucl.\ Phys.\ Proc.\ Suppl.\  {\bf 141}, 22 (2005).
%%CITATION = HEP-LAT 0501004;%% .

\bibitem{ref:Y-DGL}
S.~Kamizawa, Y.~Matsubara, H.~Shiba and T.~Suzuki,
  %``A Static baryon in a dual Abelian effective theory of QCD,''
  Nucl.\ Phys.\ B {\bf 389}, 563 (1993);
%%CITATION = NUPHA,B389,563;%%
  M.~N.~Chernodub and D.~A.~Komarov,
  %``String representation of SU(3) gluodynamics in the abelian projection,''
  JETP Lett.\  {\bf 68}, 117 (1998).
%%CITATION = HEP-TH 9809183;%%

\bibitem{ref:DualSuperconductor}
Y.~Nambu,
  %``Strings, Monopoles, And Gauge Fields,''
  Phys.\ Rev.\ D {\bf 10}, 4262 (1974);
%%CITATION = PHRVA,D10,4262;%%
G.~'t~Hooft, in {\it High Energy Physics}, ed. A. Zichichi,
EPS International Conference, Palermo (1975);
S.~Mandelstam, {Phys.\ Rept.}  {\bf 23}, 245 (1976).
%%CITATION = PRPLC,23,245;%%

\bibitem{ref:Reviews}
T.~Suzuki, Nucl.\ Phys.\ Proc.\ Suppl.\  {\bf 30}, 176 (1993);
%%CITATION = NUPHZ,30,176;%%
M.~N.~Chernodub, M.~I.~Polikarpov,
%``Abelian projections and monopoles'',
in "Confinement, duality, and nonperturbative aspects of QCD",
Ed. by P. van Baal, Plenum Press, p. 387, hep-th/9710205 (1997);
%%CITATION = HEP-TH 9710205;%%
R.W. Haymaker, Phys.\ Rept.\  {\bf 315}, 153 (1999).
%%CITATION = HEP-LAT 9809094;%%

\bibitem{Kuzmenko:2003ck}
D.~S.~Kuzmenko and Y.~A.~Simonov,
%``Triangular and Y-shaped hadrons with static sources,''
Phys.\ Atom.\ Nucl.\  {\bf 67}, 543 (2004)
[Yad.\ Fiz.\  {\bf 67}, 561 (2004)].
%%CITATION = HEP-PH 0302071;%%

\bibitem{ref:Helsinki}
A.~M.~Green, C.~Michael and M.~E.~Sainio,
  %``Four quark binding energies from SU(2) lattice Monte Carlo,''
  Z.\ Phys.\ C {\bf 67}, 291 (1995).
  %%CITATION = HEP-LAT 9404004;%%
A.~M.~Green, J.~Lukkarinen, P.~Pennanen and C.~Michael,
  %``A Study of degenerate four quark states in SU(2) lattice Monte Carlo,''
  Phys.\ Rev.\ D {\bf 53}, 261 (1996);
  %%CITATION = HEP-LAT 9508002;%%
P.~Pennanen, A.~M.~Green and C.~Michael,
  %``Four-quark flux distribution and binding in lattice SU(2),''
  Phys.\ Rev.\ D {\bf 59}, 014504 (1999).
  %%CITATION = HEP-LAT 9804004;%%

\bibitem{ref:suganuma:4Q}
F.~Okiharu, H.~Suganuma and T.~T.~Takahashi,
{\it ``Tetraquark and pentaquark systems in lattice QCD''},
hep-ph/0507187.
%%CITATION = HEP-LAT 0507187;%%

\bibitem{ref:alexandrou:4Q}
  C.~Alexandrou and G.~Koutsou,
  %``The static tetraquark and pentaquark potentials,''
  Phys.\ Rev.\ D {\bf 71}, 014504 (2005).
%%CITATION = HEP-LAT 0407005;%%

\bibitem{ref:suganuma:5Q}
F.~Okiharu, H.~Suganuma and T.~T.~Takahashi,
  %``First study for the pentaquark potential in SU(3) lattice QCD,''
  Phys.\ Rev.\ Lett.\  {\bf 94}, 192001 (2005).
  %%CITATION = HEP-LAT 0407001;%%

\bibitem{ref:alexandrou:5Q}
  C.~Alexandrou, G.~Koutsou and A.~Tsapalis,
  %``The pentaquark potential, mass and density-density correlator,''
  Nucl.\ Phys.\ Proc.\ Suppl.\  {\bf 140}, 275 (2005).
%%CITATION = HEP-LAT 0409065;%%

\bibitem{ref:Suzuki:intermeson}
  H.~Kodama, Y.~Matsubara, S.~Ohno and T.~Suzuki,
  %``Inter-meson potentials in dual Ginzburg-Landau theory,''
  Prog.\ Theor.\ Phys.\  {\bf 98}, 1345 (1997).
  %%CITATION = HEP-PH 9704340;%%

\bibitem{ref:SU3:string:interaction}
D.~Antonov and D.~Ebert,
%``String representation of field correlators in the SU(3)-gluodynamics,''
Phys.\ Lett.\ B {\bf 444}, 208 (1998);
%%CITATION = HEP-TH 9809018;%%
M.~N.~Chernodub,
%``Classical string solutions in effective infrared theory of SU(3) gluodynamics,''
Phys.\ Lett.\ B {\bf 474}, 73 (2000).
%%CITATION = HEP-PH 9910290;%%

\bibitem{ref:Simonov:Shevchenko}
V.~I.~Shevchenko and Y.~A.~Simonov,
  %``Interaction of Wilson loops in confining vacuum,''
  Phys.\ Rev.\ D {\bf 66}, 056012 (2002).
  %%CITATION = HEP-PH 0204285;%%

\bibitem{ref:X}
Belle Collaboration (S.~K.~Choi {\it et al.}), { Phys. Rev. Lett }, {\bf 91}, 262001 (2003);
CDF II Collaboration (D.~Acosta {\it et al.}), {\it ibid.} {\bf 93}, 072001 (2004);
D0 Collaboration (V.~M.~Abazov {\it et al.}), {\it ibid.} 162002 (2004);
BABAR Collaboration (B.~Aubert {\it et al.}), {\it ibid.} 041801 (2004).

\bibitem{ref:X:theory}
F.~E.~Close and S.~Godfrey,
  %``Charmonium hybrid production in exclusive B meson decays,''
Phys.\ Lett.\ B {\bf 574}, 210 (2003);
  %%CITATION = HEP-PH 0305285;%%
F.~E.~Close and P.~R.~Page,
  %``The D*0 anti-D0 threshold resonance,''
  Phys.\ Lett.\ B {\bf 578}, 119 (2004).
%%CITATION = HEP-PH 0309253;%%

\bibitem{ref:Ds}
BABAR Collaboration (B.~Aubert {\it et al.}), Phys. Rev. Lett. {\bf 90}, 242001 (2003);
Belle Collaboration (P.~Krokovny {\it et al.}), {\it ibid.} {\bf 91}, 262002 (2003).

\bibitem{hyp2}
A.~Hasenfratz, R.~Hoffmann and F.~Knechtli,
%``The static potential with hypercubic blocking,''
Nucl.\ Phys.\ Proc.\ Suppl.\  {\bf 106}, 418 (2002).

\bibitem{ref:Morningstar}
K.~J.~Juge, J.~Kuti and C.~Morningstar,
%``Fine structure of the QCD string spectrum,''
Phys.\ Rev.\ Lett.\  {\bf 90}, 161601 (2003).
%%CITATION = HEP-LAT 0207004;%%

\bibitem{ref:gluonic:excitation}
T.~T.~Takahashi and H.~Suganuma,
  %``The gluonic excitation of the three-quark system in SU(3) lattice QCD,''
  Phys.\ Rev.\ Lett.\  {\bf 90}, 182001 (2003);
  %%CITATION = HEP-LAT 0210024;%%
Phys.\ Rev.\ D {\bf 70}, 074506 (2004).
  %%CITATION = HEP-LAT 0409105;%%

\bibitem{ref:AbelianDominance}
T.~Suzuki and I.~Yotsuyanagi, { Phys.\ Rev.} {\bf D42}, 4257 (1990);
%%CITATION = PHRVA,D42,4257;%%
G.~S. Bali, V.~Bornyakov, M.~M\"uller-Preussker and K.~Schilling,
{Phys.\ Rev.}  {\bf D54}, 2863 (1996).
%%CITATION = HEP-LAT 9603012;%%

\bibitem{MILC}
C.~W.~Bernard {\it et al.},
%``The QCD spectrum with three quark flavors,''
Phys.\ Rev.\ D {\bf 64} (2001) 054506

\bibitem{GC}
The Gauge Connection {\tt http://qcd.nersc.gov/}

\bibitem{ape}
M.~Albanese {\it et al.}  [APE Collaboration],
%``Glueball Masses And String Tension In Lattice QCD,''
Phys.\ Lett.\ B {\bf 192}, 163 (1987).

\bibitem{smearparm}
C.~Legeland, B.~Beinlich, M.~Lutgemeier, A.~Peikert and T.~Scheideler,
%``The string tension in SU(N) gauge theory from a careful analysis of
%smearing parameters,''
Nucl.\ Phys.\ Proc.\ Suppl.\  {\bf 63} (1998) 260

\bibitem{hyp1}
A.~Hasenfratz and F.~Knechtli,
%``Flavor symmetry and the static potential with hypercubic blocking,''
Phys.\ Rev.\ D {\bf 64}, 034504 (2001).

%-------------------------------------------------------------------------------------------
\end{thebibliography}
